\documentclass[12pt, a4paper]{article}
\usepackage[utf8]{inputenc}
\usepackage[T1]{fontenc}
\usepackage{graphicx}
\usepackage{authblk}
\usepackage{amsmath}
\usepackage{xcolor}
\usepackage{lineno}
\usepackage{setspace}
\usepackage{hyperref}
\usepackage{amssymb}
\usepackage{xurl}

\usepackage{parskip}

\usepackage[font=footnotesize]{caption}

\usepackage{geometry}
\newgeometry{vmargin={15mm}, hmargin={20mm,20mm}}

\makeatletter
\newcommand*{\centerfloat}{%
  \parindent \z@
  \leftskip \z@ \@plus 1fil \@minus \textwidth
  \rightskip\leftskip
  \parfillskip \z@skip}
\makeatother

\usepackage{natbib}
\usepackage{lipsum}
\title{Cell differentiation can underpin the reproducibility of morphogenesis}

\author[1,2*]{Dominic Devlin}
\author[1]{Austen RD Ganley}
\author[1,3,4]{Nobuto Takeuchi}
\affil[1]{School of Biological Sciences, University of Auckland, Private Bag 92019, Auckland 1142, New Zealand}
\affil[2]{Graduate School of Arts and Sciences, The University of Tokyo, Komaba 3-8-1, Meguro-ku, Tokyo 153-8902, Japan}
\affil[3]{Research Centre for Complex Systems Biology, Universal Biology Institute, University of Tokyo, Komaba 3-8-1, Meguro-ku, Tokyo 153-8902, Japan}
\affil[4]{Department of Biology, Faculty of Sciences, Kyushu University, Fukuoka, Japan}
\affil[*]{dominicdevlin@g.ecc.u-tokyo.ac.jp}
\date{}

\begin{document}
\maketitle
\renewcommand{\abstractname}{Summary}
\begin{abstract}

Morphogenesis of complex body shapes is reproducible despite the noise inherent in the underlying morphogenetic processes. However, how these morphogenetic processes work together to achieve this reproducibility remains unclear. Here, we ask how morphogenetic reproducibility is realised by developing a computational model that evolves complex morphologies. We find that evolved, complex morphologies are reproducible in a sizeable fraction of simulations, despite no direct selection for reproducibility. We show that high reproducibility is caused by segregating moving cells that ``shape'' morphologies from stationary cells that ``maintain'' morphologies during morphogenesis. Strikingly, most highly reproducible morphologies also evolved cell differentiation, where proliferative, moving stem cells (i.e., progenitor cells) irreversibly differentiate into non-dividing, stationary differentiated cells. These results suggest that cell differentiation observed in natural development plays a fundamental role in morphogenesis in addition to the production of specialised cell types. This previously-unrecognised role of cell differentiation has major implications for our understanding of how morphologies are generated and regenerated.

\end{abstract}

\section{Introduction}

Morphogenesis is the multifaceted process that transforms relatively homogeneous starting materials---such as a zygote or fields of progenitor cells---into the complex morphological structures, such as organs, tissues and appendages, that constitute a mature organism \cite{hogan1999morphogenesis, wolpert2015principles}. This transformation occurs through a combination of chemical-level pattern formation and cellular-level shape formation \cite{goodwin1993morphogenesis, kauffman1993origins}. At the chemical level, reacting and diffusing chemicals produce spatial patterns, such as stripes and segments \cite{tautz2004segmentation, staps2023development}. At the cellular-level, processes such as cell motion, division, contraction and differential adhesion interact with chemical-level pattern formation to produce morphological shapes, such as tails, tubes, branches and limbs \cite{hogan1999morphogenesis, wolpert2015principles, zuniga2015next, iruela2013tubulogenesis, miao2023reconstruction}. For extant animals, the morphological structures produced by these processes are not only complex, but also reproduced across generations. Understanding how complex morphogenesis is made reproducible has intrigued the minds of thinkers since Pythagoras and Aristotle \cite{aristotle}, and is a focal point of developmental biology \cite{kauffman1993origins, felix2015pervasive, green2017developmental, osterwalder2018enhancer}.  

The reproducibility of morphogenesis requires both cell-level and chemical-level processes to be robust to noise \cite{swain2002intrinsic, tsimring2014noise}. Much attention has been devoted to understanding the robustness of chemical-level pattern formation to molecular sources of noise, such as fluctuations in chemical concentrations \cite{briscoe2015morphogen}, resulting in the characterisation of chemical-level processes that enhance pattern reproducibility, such as genetic feedback loops and signal transduction pathways \cite{crampin1999reaction, freeman2000feedback, eldar2002robustness, kitano2004biological, rogers2011morphogen, madamanchi2021diversity, reinitz2023robust, majka2024pattern}. In contrast, much less is known about the cellular processes underlying the robustness of morphogenesis to cell-level sources of noise, such as stochasticity in the motion and geometry of cells \cite{gilmour2017morphogen}. To address this issue, previous studies have taken a targeted approach in which they selected a set of cell-level processes, such as cell-cell signalling and adhesion between cell types, and examined each process to determine whether it increases morphogenetic reproducibility \cite{hagolani2019cell, cano2023origins}.

Here, we instead asked whether morphogenetic reproducibility is an emergent by-product of morphogenesis that evolves even if reproducibility is not explicitly selected for, and, if so, how this reproducibility is realised---a non-prescriptive approach developed by Hogeweg, Ten Tusscher \& Vroomans \cite{hogeweg1998searching, hogeweg2000evolving, hogeweg2000shapes, ten2011evolution, vroomans2016silico, vroomans2018around}. To determine whether reproducibility is a by-product of morphogenesis, we computationally generated an ensemble of ``morphogeneses'' by repeatedly evolving a population of morphologies selected for geometrically complex, multicellular shapes. We found that a sizeable fraction of evolved morphologies had high reproducibility, even though reproducibility was not explicitly selected. Strikingly, these morphologies shared one cell-level feature responsible for morphogenetic reproducibility: a ``morphogenetic division of labour'', where moving and dividing progenitor cells ``shape'' morphologies, while non-moving and non-dividing differentiated cells spatially ``anchor'' this shaping process, thereby enhancing morphogenetic reproducibility.





\section{Results}
\subsection{Multi-scale model}

\begin{figure}[t!]
    \centering
    \includegraphics{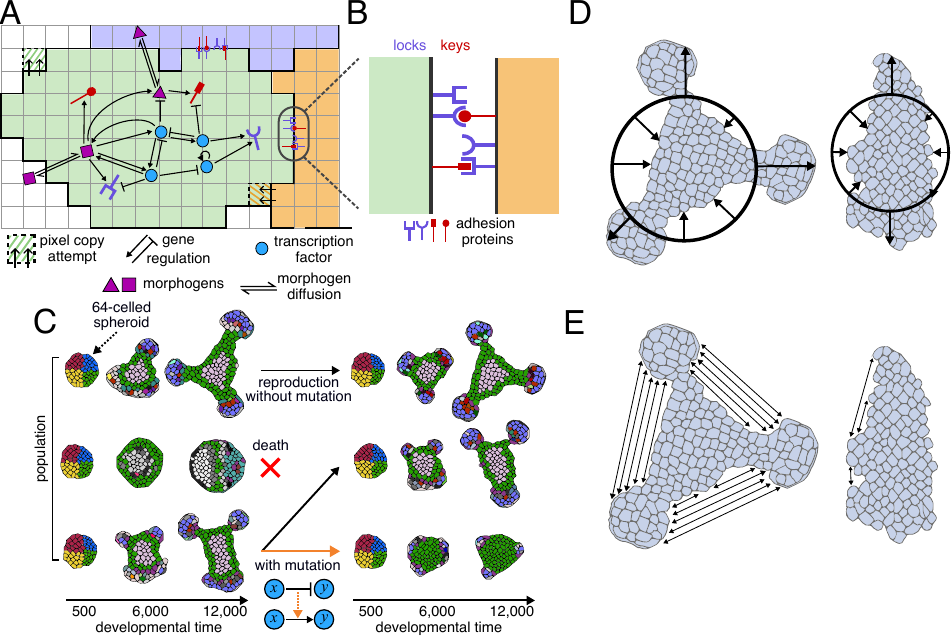}
    \caption{\textbf{Multi-scale model of morphogenesis evolution. A} Three neighbouring cells on a CPM grid. A cell consists of one or more pixels. Each cell is coloured by its ``cell state'' defined by the concentrations of all its proteins converted to boolean values (the medium is represented by white pixels). Pixels with alternating stripes indicate pixel copies at cell boundaries. Each cell contains a genome that encodes transcription factors (TFs; circles, squares, triangles), adhesion proteins (sticks with a lock or key) and membrane tension proteins (not shown). Arrows indicate regulation of gene expression by TFs (arrow head for activation, blunt head for inhibition). Double harpoon arrows indicate diffusion of morphogens (membrane-permeable TFs). \textbf{B} Adhesion proteins facilitate the binding of cells to each other via a lock and key mechanism, or to the surrounding medium (not shown). \textbf{C}. A population consists of 60 morphologies (only three depicted). Morphologies undergo a developmental phase on separate CPM grids for 12,000 DTS, and then a reproduction phase, where morphologies with complex shapes are selected. Reproduction can occur without mutation (black arrows) or with mutation (orange arrow), with mutation determined probabilistically. Mutations change the topology of the GRN (dashed orange arrow), with the example showing a change from inhibition of gene $y$ by gene $x$ to activation of gene $y$ by gene $x$. \textbf{DE} Illustration of the two measurements used to determine the complexity score on two morphologies (described in Methods \ref{evolution}). (D) depicts the measurement of how much the morphology deviates from a perfect circle (black), with the centre of the circle being the morphology's centre of mass. Black arrows at equidistant intervals around the circle mark directions where the morphology's shape deviates from the circle, with longer arrows contributing to a higher score. (E) depicts the measurement of inward folding using double sided arrows, with more arrows contributing to a higher score. The cells are all coloured grey to emphasise that only the shape, not cell states, determine the complexity score.}
    \label{model}
\end{figure}

We set out to investigate whether complex, evolvable morphogenesis is intrinsically reproducible by constructing a deliberately simplified model of real morphogenesis. To capture the noisy dynamics of real morphogenesis, we employed the Cellular Potts Model (CPM), which uses a Metropolis algorithm to simulate stochastic cell motion and cell shape dynamics (Methods \ref{cpm}) \cite{graner1992simulation, hirashima2017cellular}. The CPM models the development of a group of cells on a two-dimensional square grid ($250\times 250$ pixels). Each group of cells on the CPM grid represents a segment of developing tissue that we term a ``morphology''. Each cell of a morphology consists of a collection of neighbouring pixels on the grid (Fig.\,\ref{model}A). Pixels not occupied by cells represent the medium (Fig.\,\ref{model}A), which is akin to an extracellular matrix or fluid \cite{rozario2010extracellular}. Cell motion occurs through stochastic extensions and retractions of cell boundaries that are generated by pixel copying at these boundaries (Fig.\,\ref{model}A; Methods \ref{cpm}). Pixels on the grid are chosen in a random order with replacement for copy attempts. The unit of time is the number of pixel copy attempts equal to the total number of pixels on the grid, hereafter referred to as a developmental time step (DTS). The probability that a pixel copy occurs is determined by the minimisation of free energy arising from cell-cell adhesion, cell-medium adhesion, cell shape and cell size, as described later. 

To simulate the development of a morphology, the CPM is run for 12,000 DTS, which is approximately the minimum time it takes for a fast-growing morphology to reach the edge of the grid. Each morphology starts as a ``spheroid'' shape consisting of 64 cells of approximately 75 pixels each (Fig.\,\ref{model}C, Methods \ref{development}). After the spheroid is initialised, cells grow if they are mechanically stretched and shrink if they are mechanically squeezed (Methods \ref{development}). Cell stretching and squeezing are induced by adhesion to neighbouring cells and the extracellular medium, although it can also occur stochastically. When a cell reaches a size of 100 pixels, it divides by splitting along its minor axis into two daughter cells of approximately equal size. Linking cell division to stretching is a simplified mechanism of growth reflecting the mechano-sensitive cell division observed in several developmental contexts \cite{chen1997geometric, nelson2005emergent, guillot2013mechanics}.


To model basic cell mechanics, we equip cells with genes encoding three kinds of proteins that are near-universally present in animal cells during development: adhesion proteins, membrane tension proteins and transcription factors that regulate gene expression. Adhesion proteins modulate the adhesiveness of cells to neighbouring cells and the extracellular medium (Fig.\,\ref{model}AB). Adhesion proteins are modelled as either locks or keys, where the adhesion energy between two neighbouring cells is proportional to the number of complementary lock-key pairs they express. Similarly, the adhesion energy of a cell to the medium scales with the number of medium-adhesion proteins expressed by that cell (Methods \ref{proteins}). Directed cell motion in our model occurs through differential adhesion. Differential adhesion is the rearrangement of cells to maximise contact with surfaces they adhere to strongly and minimise contact with those they adhere to weakly. These surfaces can be other cells or the surrounding medium. Differential adhesion drives several processes essential for real morphogenesis, such as cell sorting, formation of tissue boundaries and cell migration \cite{wolpert2015principles, steinberg2007differential}. Adhesion proteins also influence cell fluidity in our model, i.e., the rate of cell rearrangements. When the free energy arising from cell-cell adhesion is low, cells behave more fluid-like (e.g., mesenchymal cells). When this energy is high, cells behave more solid-like (e.g., epithelial cells) \cite{chiang2016glass}.

Membrane tension proteins restrict cell motion by making the cell shape less deformable by energetically constraining the length of its dynamically determined longest axis, irrespective of the cell's orientation. The constraint on the length of the longest axis decreases with the number of expressed tension proteins (Methods \ref{proteins}). These proteins model an increase in cell membrane tension, which occurs in real cells by the accumulation of actin filament stress \cite{sehring2014equatorial}. To maintain simplicity and broad applicability, we do not implement polarised cell tension or contractility, properties required for processes like invagination.

To model gene regulation, we couple each CPM cell to gene expression dynamics, as previously done \cite{hogeweg2000evolving, hogeweg2000shapes, knabe2008evolution, colizzi2020evolution, hirway2021multicellular, vroomans2023evolution}. Each cell is equipped with transcription factors (TFs), that form a gene regulatory network (GRN) with its adhesion and membrane tension proteins. A GRN is a graph consisting of nodes representing proteins and edges representing TF-mediated activation or inhibition of gene expression (Fig.\,\ref{model}A). Concentrations of proteins within a cell are determined by numerically integrating a set of ordinary differential equations given by the GRN (Methods \ref{grn}). To model cell-cell signalling, a minority fraction of TFs (hereafter called morphogens) diffuse between cells and into the medium (Fig.\,\ref{model}A; Methods \ref{grn}). These morphogens model those encountered in real development, such as Wnts and BMPs \cite{eldar2002robustness, martin2012canonical, farin2016visualization}. Morphogens allow different cells to express different proteins even though all cells within a morphology have identical genes and GRNs. 

The other way differential protein expression occurs in natural development is by the asymmetric distribution of proteins between cells in a tissue \cite{zhang2018symmetry}. To model this, we distribute the concentrations of two non-morphogen TFs asymmetrically in the 64-celled spheroids. We restrict one of these TFs to 32 cells left of the vertical centre line and the other to 32 cells below the morphology's horizontal centre line. Thus, the cells in each of the four sectors in the spheroid each have unique concentrations of these two TFs (illustrated in Fig.\,\ref{model}C), although whether this asymmetry persists over development depends on the gene regulatory network.


We distinguish cells based on the proteins they express by assigning each cell a cell state, defined as a vector of boolean values, where each boolean value indicates whether a tension or adhesion protein is expressed or not expressed (Methods \ref{statespace}). All cells with the same state are shown with the same colour on the CPM grid (Fig.\,\ref{model}). Cell states are only used to visualise and analyse model outcomes and do not play any role in model dynamics. 

To simulate the evolution of morphogenesis, we established an initial population of 60 morphologies (Fig.\,\ref{model}C; Methods \ref{evolution}), with each assigned a different randomly generated GRN. Each morphology develops on a separate CPM grid. We applied a genetic algorithm to select for shape complexity by measuring inward folding and deviation of the morphological shape from a circle at the end of the 12,000 DTS (Fig.\,\ref{model}DE; Methods \ref{evolution}). The 15 morphologies with the highest shape complexity reproduce four times to populate the next generation, and their GRNs undergo a single mutation with a probability of 50\%. A mutation changes the regulatory effect of one TF on one gene, such as causing a TF to switch from inhibiting the expression of a gene to activating its expression (see Fig.\,\ref{model}C). The GRNs used in our main set of simulations are comprised of nine transcription factors (including three morphogens), 15 adhesion proteins and two membrane tension proteins. Gene duplication and deletion do not occur. To broadly explore the types of morphogenesis our model evolves, we also run simulations where morphologies start as a rectangular initial shape with proteins asymmetrically distributed along the longest axis of the rectangle (Fig.\,S9), simulations where morphogen diffusivity mutates alongside GRN mutations (Fig.\,S9), simulations with different selection pressures (Fig.\,S10) and simulations with other numbers of genes (Fig.\,S11). 




\subsection{Reproducibility is not an intrinsic property of complex morphogenesis}

\begin{figure}
    \centerfloat
    \includegraphics{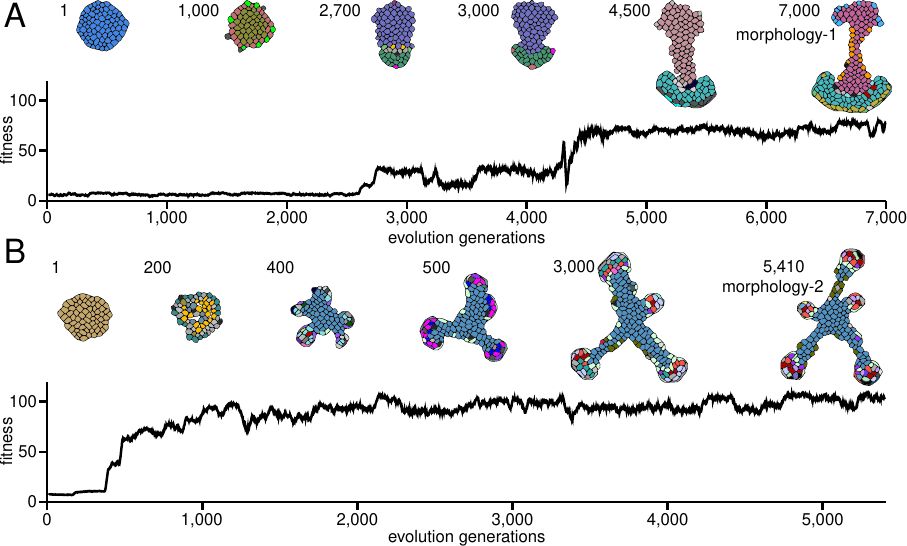}
    \caption{\textbf{Computational simulations showing the evolution of complex morphologies. AB} The plots show a five-generation moving average of population fitness (shape complexity) over evolution. The morphologies above the plots each depict a separate morphology at the end of its development (12,000 DTS) from six different generations over the course of evolution. The generation number of each morphology is shown to its left. ``Evolved morphologies'', defined as the most complex morphology from the final generation of the simulation, are shown on the far right. The evolved morphology in (A) and (B) are referred to as morphology-1 and morphology-2, respectively.} 
    \label{fig1}
\end{figure}

To investigate whether morphogenesis is intrinsically reproducible, we conducted 126 independent evolutionary simulations of our model starting from a circular initial condition. We ran each simulation for at least $2.5\times10^3$ generations, which is usually sufficient to reach a plateau in fitness (Fig.\,S1). We then identified the fittest morphology (i.e., most complex shape) from the final generation (hereafter referred to as an ``evolved morphology'') from each simulation. To ensure that we were analysing the reproducibility of complex shapes, we removed 36 evolved morphologies that did not reach an arbitrary threshold of complexity (our results are similar when a different threshold is used; Fig.\,S2A). The 90 morphologies above this threshold each display a different shape (Fig.\,S1), with the evolution of two representative morphologies illustrated in Fig.\,\ref{fig1}A and B, termed morphology-1 and morphology-2, respectively. 

\begin{figure}[t!]
    \centerfloat
    \includegraphics{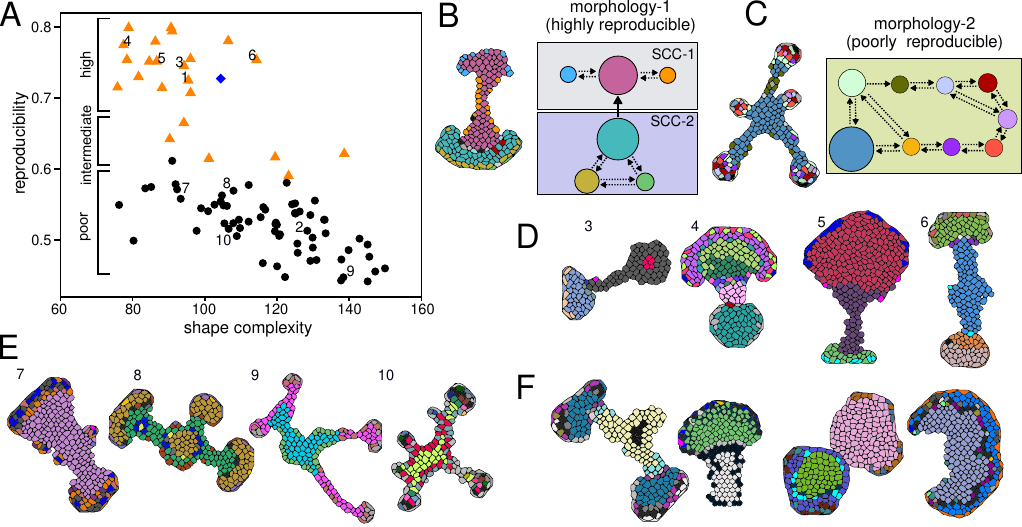}
    \caption{\textbf{Both highly and poorly reproducible morphologies evolve in response to selection for shape complexity. A} Reproducibility scores against shape complexity scores for the 90 morphologies that reached a threshold of shape complexity. Circles indicate morphologies with a single strongly connected component (SCC, defined in Section~\ref{stem-section}) (mean reproducibility$=52.0\%$, $n=65$); filled triangles indicate morphologies with multiple SCCs with unidirectional transitions between them (mean reproducibility$=72.1\%$, $n=24$). The blue diamond is a morphology with multiple SCCs without unidirectional transitions (Fig.\,S8A). Numbered arrows refer to the morphologies in panels B, C, D, and E. \textbf{BC} Simplified cell state spaces for (B) morphology-1 and (C) morphology-2. Node colours correspond to cell state colours depicted in the morphologies to the left each of cell state space. Arrows are cell-state transitions. Cell states are partitioned into strongly connected components (SCCs, coloured boxes). Node sizes depict cell state frequency over all of development; node colours correspond to cell states from morphology-1 and morphology-2, respectively (shown to the left of each state space). See Fig.\,S5IJ for the state spaces without pruning of nodes and edges. \textbf{D} Four morphologies that are highly reproducible. \textbf{E} Four morphologies that are poorly reproducible. \textbf{F} Four highly reproducible morphologies from simulations where morphologies evolved with morphogens mutating (for more information see Fig.\,S9).}
    \label{results}
\end{figure}


We examined whether the 90 evolved morphologies with complex shapes display reproducible morphogenesis by repeatedly simulating their development with different pixel copy orders (60 replicates per morphology, hereafter referred to as developmental replicates). To quantify reproducibility, we measured a ``reproducibility score'', which indicates how geometrically similar the morphologies of replicates are to each other (Methods \ref{measurements}). The reproducibility score depends on the geometry, size and time taken to generate the morphology but is invariant to reflection, rotation and translation of the morphology. The distribution of reproducibility scores across all 90 evolved morphologies is bimodal (bimodality coefficient $=0.69$, Fig.\,\ref{results}A), with 19 morphologies in the upper mode, which we term highly reproducible (including morphology-1; Fig.\,\ref{results}D shows four other examples), 65 morphologies in the lower mode, which we term poorly reproducible (including morphology-2; Fig.\,\ref{results}E shows four other examples), and six that are between the two modes, which we term intermediately reproducible (see Fig.\,S6 for information about intermediately reproducible morphologies). The bimodal distribution implies that evolved morphologies consist of a mixture of two populations with distinct properties.

\subsection{Highly and poorly reproducible morphologies have distinct cell-state transition dynamics}

We investigated whether differences in reproducibility trivially arise from variation in shape complexity (Text S1, Fig.\,S2D-G) or variation in the number of cell states observed over development (Fig.\,S3). We found that differences in reproducibility between morphologies categorised as highly and poorly reproducible could not be explained exclusively by shape complexity or number of cell states. Although variation in the number of cell states does not explain reproducibility differences, we noticed a stark difference in the spatial distribution of cell states between highly and poorly reproducible morphologies (as depicted by the distribution of cell colours in Figs.\,\ref{results}B-F). Specifically, each cell state localises to a ``domain'' of highly reproducible morphologies, whereas each cell state reappears all over poorly reproducible morphologies. The spatial distribution of cell states is determined by how and which cells transition between states. Thus, we wondered whether highly and poorly reproducible morphologies have different cell-state transition dynamics. To this end, we recorded all cell-state transitions of all cells in each evolved morphology to generate a ``cell state space'' for each morphology, which is defined as a graph consisting of nodes representing cell states and edges representing all possible transitions between cell states (see Methods \ref{statespace} for details). Since each cell state space contains numerous nodes and edges, we simplified the state spaces in two steps. First, we pruned infrequently observed cell states and cell-state transitions (Methods \ref{statespace}). Second, we split the cell state space into strongly connected components (SCCs), where an SCC is defined as a set of cell states for which a pathway of transitions exists from any cell state to any other cell state within that set. We found that almost all (64 out of 65) poorly reproducible morphologies had just a single SCC, as illustrated for morphology-2 in Fig.\,\ref{results}C. In contrast, all 19 highly reproducible morphologies and most (five out of six) intermediately reproducible morphologies had cell state spaces containing multiple SCCs, as illustrated for morphology-1 in Fig.\,\ref{results}B. Moreover, most (24 out of the 25) morphologies with multiple SCCs had at least one SCC that unidirectionally transitioned to another SCC (Fig.\,\ref{results}A; see Fig.\,S8AB for details about morphologies without unidirectional SCC transitions). The difference in reproducibility between morphologies with a single SCC and those with multiple SCCs is statistically significant ($p<10^{-12}$, two-tailed t-test), suggesting that the presence of multiple SCCs is involved in morphogenetic reproducibility.



\begin{figure}[t!]
    \centerfloat
    \includegraphics{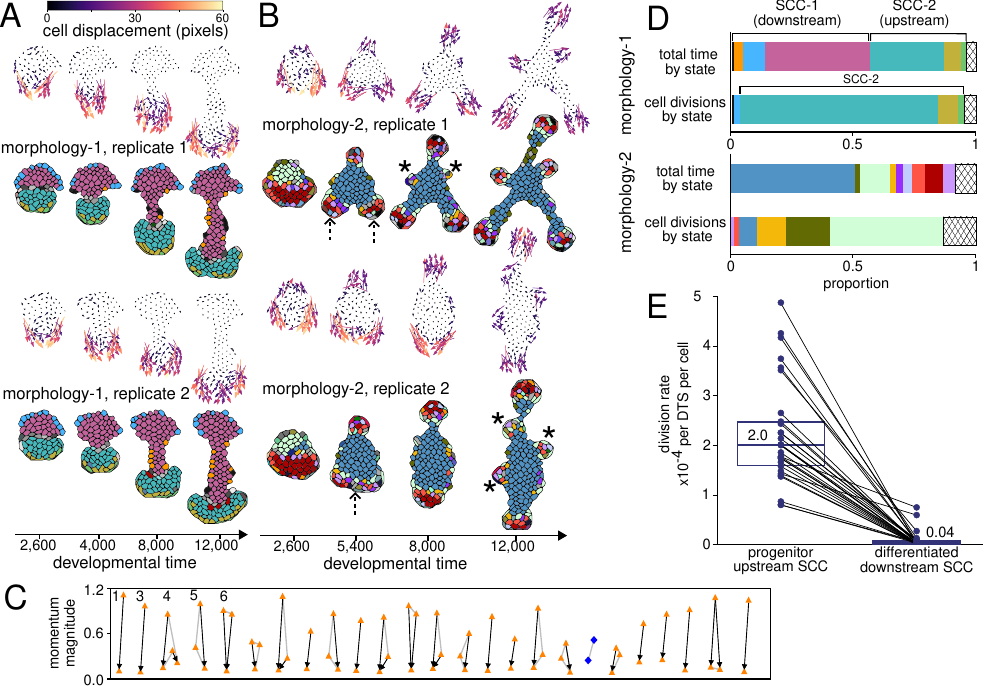}
    \caption{\textbf{Highly reproducible morphologies have moving and dividing cells that undergo unidirectional transitions to non-moving and non-dividing cells. AB} Two developmental replicates of morphology-1 (A) and morphology-2 (B) are depicted after 2,600, 4,000, 8,000 and 12,000 DTS, showing a difference in their reproducibility. Dashed arrows in (B) indicate the presence (replicate-1) or absence (replicate-2) of a bifurcation in collective cell motion; asterisks indicate protrusions. Vector plots show the displacement of the centre of mass of each cell during 2,000 DTS at each respective time point, with colours indicating magnitude (the lighter, the larger). \textbf{C} Average cell momentum magnitude for each SCC from the 25 morphologies with multiple SCCs. Momentum is the distance travelled by a cell per DTS multiplied by its size in pixels (see Methods \ref{Momentum}). Black arrows indicate unidirectional transitions between SCCs. Grey lines connect SCCs from the same morphology that do not have unidirectional SCC transitions between each other. Filled orange triangles are SCCs from morphologies that have unidirectional SCC transitions. All transitory SCCs are excluded (see Methods \ref{statespace} for information about transitory SCCs). Blue diamonds are SCCs from the highly reproducible morphology that does not have unidirectional SCC transitions. The numbered morphologies correspond to those from Fig.\,\ref{results}D. \textbf{D} Stacked bar charts showing the proportion of developmental time spent in each cell state and the proportion of cell divisions undergone by each state across all cells during developmental replicate-1 of morphology-1 and development replicate-1 of morphology-2. Diagonal lattices are pruned states. \textbf{E} The rate at which cells divide per developmental time when their state belongs to an upstream SCC (left) or a downstream SCC (right). Each data point represents an SCC from (C). Black lines connect upstream SCCs to their counterpart downstream SCCs. Boxes show medians and interquartile ranges (IQR); the downstream SCC box is tiny because most division rates are either very low or 0. Numbers on top of the box plots are median cell division rates.}
    \label{fig3}
\end{figure}

To understand why the number of SCCs affects morphogenetic reproducibility, we examined the motion of cells and their states, given that cell motion ultimately determines the morphology. To investigate this, we visualised the velocity of every cell at multiple time points during the development of evolved morphologies (here using morphology-1 and morphology-2 as examples). We found that in morphology-1, which is highly reproducible, cells in states belonging to one of its two SCCs collectively move downwards and radiate slightly outwards, like a travelling wave (Fig.\,\ref{fig3}A; Movie S1), while cells in states belonging to the other SCC show little motion. These position-dependent collective cell motions generate a ``cap'' of moving cells and a ``stalk'' of stationary cells forming in the wake of the moving cap (Fig.\,\ref{fig3}A), with cap cells being more fluid-like and stalk cells more solid-like due to expressing different adhesion and tension proteins (Fig.\,S7). Importantly, these collective cell motions are consistent across developmental replicates (Fig.\,\ref{fig3}A). This result indicate that stationary and moving cell states are segregated into distinct domains of morphology-1, with the cell states of each domain corresponding to distinct SCCs. Thus, transitions between moving and stationary states are constrained in space (into separate domains) and dynamically (into separate SCCs). The segregation of moving and stationary states results in high reproducibility for developmental dynamics that are different from morphology-1, such as morphologies that undergo epiboly (Fig.\,\ref{fungi}F, Movie S4), morphologies that branch (Fig.\,S8CD, Movie S5) and morphologies with no transitions between moving and stationary states (Fig.\,S8AB).

By contrast, in morphology-2 (one SCC), inconsistent cell motion across replicates coincides with inconsistent cell-state distributions (Fig.\,\ref{fig3}B; Movie S2). These inconsistencies arise via ``bifurcations'' and ``protrusions'' in morphology-2. Bifurcations are the spontaneous splitting of a cluster of moving cells, caused by some moving cells transitioning to stationary states at the bifurcation point (Fig.\,S4AB). These transitions occur by expression of proteins that increase membrane tension and adhesion energy with surrounding cells. Bifurcations occur inconsistently across developmental replicates of morphology-2 (dashed arrows, Fig.\,\ref{fig3}B). Protrusions are the spontaneous onset of collective motion when neighbouring stationary cells transition into moving cells (Fig.\,S4AC). These transitions are driven by the cyclical expression of adhesion proteins along morphogen gradients (Fig.\,S5), which causes cells to oscillate between (i) stretching outward over other cells and (ii) retracting inward. Protrusions occur inconsistently: in replicate-1, two cell groups protrude around 8,000 DTS (Fig.\,\ref{fig3}B, top), whereas in replicate-2 this occurred in three cell groups at locations different from replicate-1 around 12,000 DTS (Fig.\,\ref{fig3}B, bottom). These results indicate that moving and stationary cell states are neither separated in space nor constrained dynamically (because they are bidirectional). This lack of constraints amplifies cell-level noise to the timing and location of morphogenetic events such as bifurcations and protrusions in morphology-2, reducing reproducibility. These results also apply to different kinds of morphogenesis, such as ``finger'' formation via cell death (Fig.\,S8FG, Movie S5).



We next sought to confirm that moving and stationary SCCs are partitioned into different SCCs across all highly reproducible morphologies. To test this, we determined whether different SCCs have different cell-motion properties in morphologies with multiple SCCs by measuring the momentum of cells in each SCC (Methods \ref{Momentum}). The result shows a significant disparity (mean 6.7-fold difference) in the average magnitude of cell momentum between SCCs across all highly reproducible morphologies (Fig.\,\ref{fig3}C), indicating that moving and stationary states are indeed partitioned into separate SCCs. When there are unidirectional transitions between SCCs, the transitions are always from high momentum SCCs to low momentum SCCs (black arrows in Fig.\,\ref{fig3}C). To understand why there is a difference in cell mobility, we compared the cell-cell adhesion energy among cells from upstream SCCs to the cell-cell adhesion energy among cells from downstream SCCs, since cell-cell adhesion energies are the key component of cell mobility in our model (Methods \ref{proteins}). We found that the cell-cell adhesion energy was much lower among those from upstream SCCs (mean 1.39) than those from downstream SCCs (mean 3.51), with no overlap between groups (Fig.\,S7A). These results suggest that morphologies with multiple SCCs establish a ``morphogenetic division of labour'', whereby domains of moving cells (in upstream SCCs) ``shape'' morphologies and unidirectionally transition into domains of stationary cells (in downstream SCCs) that ``maintain'' the morphologies shaped by moving cells. Making moving-to-stationary transitions unidirectional increases reproducibility by preventing noise-induced transitions from stationary to moving cells (for instance, those that lead to protrusions in morphology-2). 



\subsection{Multiple SCCs in reproducible morphologies resemble progenitor-cell systems}
\label{stem}

Our results suggest that having a morphogenetic division of labour between SCCs is the most common way to achieve a complex yet reproducible morphology in our model. Intriguingly, these morphogenetic divisions of labour resemble ``progenitor-cell systems'', where upstream SCCs (cell states that differentiate into one or a few types, known as progenitor cells) irreversibly differentiate to downstream SCCs (differentiated cell states). Another hallmark of progenitor-cell systems is that progenitor cells undergo more rapid cell division than differentiated cells \cite{gilbert2010developmental, furusawa2001theory}. To examine whether morphologies also display this other hallmark, we compared how often cells in upstream versus downstream SCCs divide (Fig.\,\ref{fig3}DE). We found that cells in upstream SCCs divide a median of 49.6 times more frequently than those in downstream SCCs. Thus, the morphogenetic division of labour encompasses both differential cell motion and differential cell divisions between SCCs. Hereafter, we refer to upstream SCCs and downstream SCCs as progenitor-cell types and differentiated-cell types, respectively. We define a progenitor-cell system as comprising one progenitor-cell type and its counterpart differentiated-cell types (some morphologies have multiple progenitor-cell types, thus multiple progenitor-cell systems). The evolution of highly reproducible morphologies with these progenitor-cell systems also occurs for simulations with different initial shapes (Fig.\,S9A-C), simulations where morphogen diffusivity can mutate (Fig.\,S9D-F), and simulations with different numbers of genes (Fig.\,S11). Moreover, progenitor-cell systems evolve in 80-90\% of simulations when we modified the selection criterion to indirectly select morphogenesis that undergoes directional motion (which indirectly selects for reproducibility; Text S2 and Fig.\,S10), indicating that progenitor-cell systems are easy to evolve from different initial GRNs.


\subsection{Progenitor-cell systems elevate morphogenetic reproducibility by regulating cell-motion transitions at cell-type boundaries}
\label{stem-section}

\begin{figure}
    \centering
    \includegraphics{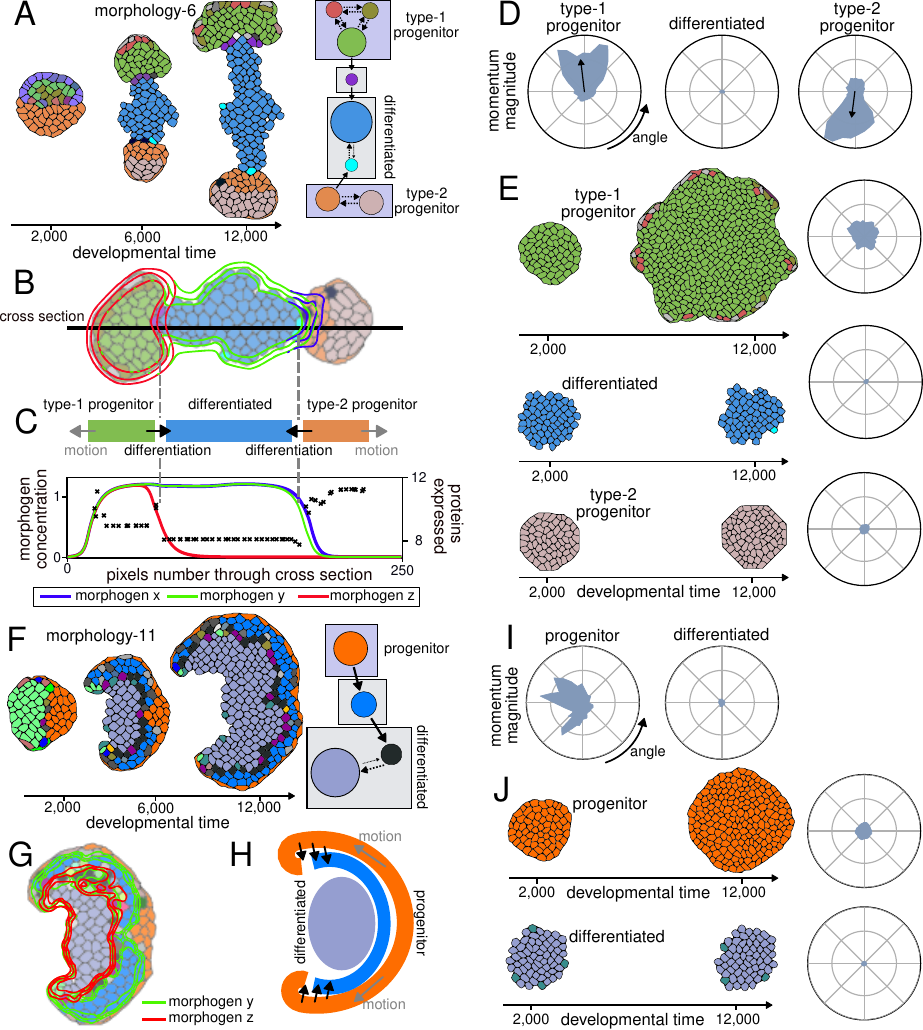}
    \caption{\textbf{Mechanisms underlying progenitor-cell differentiation and motion. A} Development and cell state space of organism-6, showing two progenitor-cell types, one differentiated-cell type, and a transitory SCC (see Methods \ref{statespace} for information about transitory SCCs). The morphology is shown after 2,000, 6,000 and 12,000 DTS. \textbf{B} Contours showing the concentrations of the three morphogens ($x$, $y$ and $z$) overlaid on morphology-6 after 9,000 DTS. Each contour joins points of equal concentration of the same morphogen. \textbf{C} Schematic depicting type-1, type-2 and differentiated cell domains from (B), with progenitor-cell motion (grey arrows) and differentiation (black arrows) indicated. Vertical dashed lines indicate cell-type boundaries. Below, morphogen concentrations along the cross-section line are plotted, along with the sums of cell protein concentrations for cells along the cross-section (each cross is one cell). \textbf{D} Polar plots of momentum magnitude by angle of motion for each cell type summed over all cells over the 12,000 DTS of morphology-6 development (Methods \ref{Momentum}). \textbf{E} Development of type-1, type-2 and differentiated cells in isolation. Polar plots show distributions of cell momentum as in (D). \textbf{F} Development and cell state space of morphology-11, showing a progenitor-cell type, a differentiated cell-type and a transitory SCC. The morphology is shown after 2,000, 6,000 and 12,000 DTS and its state space. \textbf{G} Contours showing the concentrations of morphogens $y$ (green) and $z$ (orange, morphogen $x$ is hidden for visibility) overlaid on morphology-11 after 8,000 DTS. \textbf{H} A schematic depicting progenitor and differentiated cell domains for morphology-11, with progenitor-cell motion (grey arrows) and differentiation (black arrows) indicated. The combined presence of morphogens $y$ and $z$ induces differentiation of progenitor cells. \textbf{I} Polar plots of momentum magnitude by angle of motion for progenitor and differentiated cell types summed over all cells over the 12,000 DTS of morphology-11 development shown in (F). \textbf{J} Development of progenitor and differentiated cells from morphology-11 in isolation. Polar plots show distributions of cell momentum as in (I).}
    \label{fungi}
\end{figure}

We wondered how progenitor-cell systems increase morphogenetic reproducibility despite frequent cell transitions from moving states (progenitor cells) to stationary states (differentiated cells), which cause poor reproducibility in evolved morphologies with only a single SCC. We noticed that these transitions, henceforth termed differentiation, only occur at the boundaries between domains of progenitor and differentiated cell types, suggesting that they do not impact reproducibility because they are tightly regulated in space. For example, in morphology-6, which has two progenitor cell types denoted type-1 and type-2 (Fig.\,\ref{fungi}A), progenitor cells initially differentiate exclusively at the boundary between the type-1 and type-2 progenitor cells and, subsequently, exclusively at the boundary between progenitor cells and differentiated cells (Movie S3). Both progenitor-cell domains are present from the start of morphology-6 development (i.e., the 64-celled spheroid). Whether a cell starts at type-1 or type-2 depends on the concentration of the TF distributed asymmetrically across the horizontal centre line. Since type-1 and type-2 stem cells start in distinct domains, the spatial layout of progenitor and differentiated cells ends up mirroring the topology of the cell state space, with differentiated cells consistently forming between the progenitor-cell domains (Fig.\,\ref{fungi}A; see S5K for other examples).

Boundary-localised differentiation suggests that progenitor-cell differentiation is spatially regulated. To determine how this regulation is achieved, we focused on morphology-6 as it has two separate examples of progenitor-cell differentiation. We hypothesised that boundary-localised differentiation is caused by morphogen-mediated interactions between progenitor and differentiated cells as morphogens provide a means to spatially regulate gene expression. To test this, we made a contour plot of morphogen concentrations for morphology-6 (Fig.\,\ref{fungi}B). The plot shows that the concentrations of different morphogens abruptly change at the boundaries between progenitor and differentiated cells (dashed lines in Fig.\,\ref{fungi}B). To determine whether gene expression responds to these changes in morphogen concentrations, we plotted the sum of protein concentrations (excluding morphogens) for cells along a cross-section of morphology-6 (Fig.\,\ref{fungi}C). The plot shows that this sum changes abruptly wherever morphogen concentrations change. These results suggest that a specific morphogen profile induces differentiation, with differentiated cells producing this profile and thus localising differentiation to the boundary between progenitor and differentiated cells. To directly test this, we isolated each progenitor-cell type from the other two cell types, thereby removing the effect of differentiated cells on morphogen profiles. We isolated cell types by creating morphologies in which all cells in the initial spheroid were set to the state that is most frequently observed for each of the progenitor-cell types (e.g., the green cell state shown in Fig.\,\ref{fungi}A for type-1 progenitor cells). We developed these morphologies for 12,000 DTS, and found that neither type-1 nor type-2 progenitor-cell types differentiates (Fig.\,\ref{fungi}E). To test the robustness of our results, we repeated the above analyses on morphology-11, which undergoes a different kind of morphogenesis from morphology-6---progenitor cells spread around the surface of the morphology before differentiating like epiboly \cite{bruce2016zebrafish} (Fig.\,\ref{fungi}F, Movie S4). We found that morphology-6 progenitor cells differentiate in response to the combined exposure to two types of morphogens (Fig.\,\ref{fungi}GH), and that progenitor cells do not differentiate when isolated from differentiated cells (Fig.\,\ref{fungi}J). These results support the hypothesis that differentiated cells induce progenitor-cell differentiation, thus localising differentiation to the boundaries between cell types. 


We next asked whether the above findings---that differentiated cells induce progenitor-cell differentiation---are generalisable to all morphologies with progenitor-cell systems. To answer this, we isolated each of the 30 progenitor-cell types from the 24 morphologies with progenitor-cell systems from all other cell types and tested whether the isolated progenitor cells differentiate through the same method as described in the previous paragraph. We found that the great majority (26 out of 30) do not differentiate when isolated from other cell types (Fig.\,S12 shows why the four exceptions do not counter our hypothesis). This result, along with the observation that every progenitor-cell type always differentiates at the boundary shared with differentiated cells (Figs.\,\ref{fig3}A, \ref{fungi}A-C and F-H), indicates that differentiated cells induce progenitor-cell differentiation.

\begin{figure}
    \centering
    \includegraphics{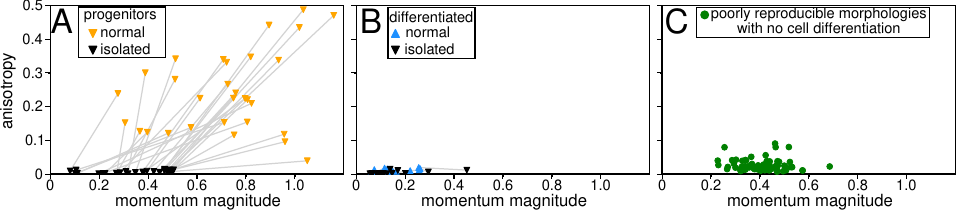}
    \caption{\textbf{Interactions between cell types drive anisotropic progenitor-cell motion. A} Motion anisotropy of progenitor cells within morphologies (orange) and isolated (black) plotted against average cell momentum magnitude ($n=30$ cell types). Grey lines connect a progenitor-cell type within a morphology to that progenitor-cell type in isolation. \textbf{B} The same as (A), but for differentiated cells ($n=26$ cell types). \textbf{C} The same as (A), but for cells from morphologies with only one SCC ($n=65$).} 
    \label{anisotropy}
\end{figure}

The spatial consistency of progenitor-cell differentiation is crucial but not sufficient for morphogenetic reproducibility. The other critical factor is the motion of progenitor cells in a consistent direction (e.g., morphology-6 Fig.\,\ref{fungi}C and morphology-11 Fig.\,\ref{fungi}H), as this motion helps to confer consistent shape formation across developmental replicates. Given that progenitor-cell differentiation is induced by differentiated cells, we hypothesised that the directionality of progenitor-cell motion is also induced by differentiated cells. To test this, we determined how progenitor cells' motion depends on the presence or absence of differentiated cells by measuring the magnitude of progenitor cells' momentum as a function of the angle of momentum, either in normal development (differentiated cells present) or when each progenitor-cell type is isolated (differentiated cells absent). The result shows that cell momentum is strongly directional (i.e., anisotropic) in normal development, whereas it is radially symmetrically distributed (i.e., isotropic) when cells are isolated (Fig.\,\ref{fungi}DEIJ shows results for morphology-6 and morphology-11; see Fig.\,S12 for other morphologies). We quantified this difference in anisotropy by calculating the ratio of the variance in momentum magnitude across angle to the mean momentum magnitude across angle (Methods \ref{Momentum}). We found that anisotropy of progenitor-cell motion decreased an average of 65-fold across the 30 progenitor-cell types when progenitor cells were isolated (Fig.\,\ref{anisotropy}). To understand mechanically how this anisotropic motion occurs, we analysed the expression of adhesion proteins across morphologies with progenitor-cell systems, given that our model is primarily adhesion-based. We consistently found differential adhesion between progenitor and differentiated cells, or differences between progenitor and differentiated cells in their adhesion to the medium across all morphologies with progenitor-cell systems (Fig.\,S7B). Together, these results suggest that interactions between progenitor and differentiated cells represent another aspect of the morphogenetic division of labour, whereby differentiated cells not only maintain the shape formed by progenitor cells but also make this shape formation consistent by inducing directional progenitor-cell motion.

\section{Discussion}

We investigated whether reproducible morphogenesis emerges in a computational model of development without being directly selected for. To search for reproducibility-conferring properties, we designed our model to simulate generic morphogenetic processes. We found that reproducible morphogenesis evolved in a minority fraction of simulations, indicating that reproducibility can be an emergent by-product of morphogenesis, but not always. Our results show two principles for reproducible morphogenesis. First, moving and stationary states should be segregated into distinct domains (e.g., tissues). Second, transitions between these states must be unidirectional and only occur at the boundary between domains. These principles are consistent with observations of real animal development, where cells in a tissue domain commit to collectively moving or remaining stationary \cite{lee2006epithelial, scarpa2016collective}, with transitions from one to the other occurring unidirectionally at tissue boundaries, e.g., the differentiation of moving mesoderm into stationary somites \cite{pourquie2001vertebrate}. Our evolutionary simulations show that the most common way to achieve highly reproducible morphogenesis is a morphogenetic division of labour based on cell differentiation, where domains of mobile, dividing progenitor cells ``shape'' morphologies and irreversibly differentiate into domains of stationary, non-dividing cells that ``maintain'' these morphologies. This labour division allows differentiated cells to serve as anchor points that establish the locations of progenitor-cell differentiation and directions of progenitor-cell motion via morphogen gradients and differential cell adhesion.  

A limitation of our current model is the diversity of morphogenesis it can evolve. Although we observed at least six kinds of morphogenesis in our model (see Figs.\,\ref{fig3}AB, \ref{fungi} and S8), this diversity is small compared to animal tissues. Diversity in our model is smaller than in animal tissues because of two aspects of our model's design. First, we implemented a set of cell properties deliberately constrained to those universally present in animal development: cell-cell adhesion, cortical tension, cell division, and gene regulation. We chose these properties to ensure the reproducibility-conferring principles we discovered would broadly apply to animal morphogenesis. However, this implementation limits the kinds of morphogenesis that can evolve to those driven by these cell properties. For instance, the absence of cell polarisation in our model is the likely reason we do not observe invaginations, since invaginations require polarisation \cite{plageman2011trio, pearl2017cellular}. Second, we only tried two evolutionary selection pressures in our model (shape complexity and directional motion). In contrast, real development is shaped by numerous and varied selection pressures acting on large populations over longer timescales. We expect that introducing more diverse selection pressures and ecologies would increase the morphological diversity our model produces. That our model does not replicate the full diversity of animal tissues does not invalidate the reproducibility-conferring processes we discovered. However, further research is needed to test whether these processes apply only to the limited kinds of morphogenesis driven by the cell properties we implemented (e.g., those involving cell divisions, differential adhesion and cortical tension) or apply to other kinds of morphogenesis.

A major finding of our results is that progenitor-cell systems may have a previously unrecognised role in achieving reproducible yet complex morphogenesis, in addition to their well-known role of generating specialised cell types \cite{wolpert2015principles, weissman2000stem, brunet2017origin, collinet2021programmed}. This proposal is supported by three key observations and their associated limitations, which we describe below.

The first observation supporting our proposal is that progenitor-cell-based morphogenesis resembles the real morphogenesis of elongated and bulge-like sub-structures. These include organ sub-structures such as intestinal crypts and villi \cite{gjorevski2011integrated, sato2013growing, huycke2024patterning}, lung alveoli \cite{rock2011epithelial, yang2014developmental}, mammary gland buds \cite{scheele2017identity}, salivary gland buds \cite{wang2021budding} and kidney tubules \cite{mcmahon2016development}. For example, in the kidney, nephron tubule formation begins with the recruitment of mesenchymal progenitor cells to a ureteric bud branch \cite{schnell2022principles}. These progenitor cells then divide and differentiate into epithelial cells to elongate the nephron, a process driven by signals that originate from already differentiated epithelial cells \cite{mcmahon2016development}, similar to the signalling between progenitor and differentiated cells that drives elongation in our evolved morphologies. Another resemblance is to gut villus elongation in vertebrates. In gut villus elongation, high surface tension, fluid-like mesenchymal tissue progressively elongates the villus before differentiating into solid-like mesenchymal tissue \cite{huycke2024patterning}. Similarly, progenitor-cell tissues in our model tend to have higher surface tension and fluidity than differentiated cells. Beyond organ sub-structures, embryonic tail elongation \cite{miao2023reconstruction} and limb elongation \cite{damon2008limb} are driven by proliferating progenitor cells at the tip of the tissue that progressively differentiate, similar to the progenitor-cell-based morphogenesis observed in our model. In contrast to our model, where morphogenesis is primarily driven by differential adhesion and cell division, the morphogenesis of some of these real sub-structures involves other dynamics such as invaginations and tubulogenesis. The fact that we did not observe these dynamics is unsurprising, because we did not include properties necessary for invaginations, such as polarised cell contractions \cite{plageman2011trio}, and we did not simulate our model in three dimensions, which is essential for tubulogenesis. 

A second observation supporting our proposal is that progenitor-cell systems can generate modular morphogenesis, a characteristic of organ development \cite{hogan1999morphogenesis}. This modularity arises from the fact that progenitor-cell-based morphogenesis depends solely on the interactions between progenitor and differentiated cells. This modularity allows progenitor-cell-based morphogenesis to occur in preconceived directions from arbitrary starting conditions (for examples of this see Fig.\,S13). Consequently, the kinds of shapes that progenitor-cell-based morphogenesis creates could constitute sub-structures within more complex morphologies, such as organ building blocks, with these sub-structures repeating wherever the appropriate cell-type configuration is present, i.e., branching morphogenesis \cite{hogan1999morphogenesis}. However, our evolved morphologies only undergo a maximum of one branching event during morphogenesis (Fig.\,S8CD, Movie S5), likely because the set of cell properties we implemented is insufficient for tissues to undergo iterative branching. We hypothesise that iterative branching could be achieved by integrating progenitor-cell-based morphogenesis with known mechanisms for branch point specification, such as localised inhibition of proliferation at the branching point (e.g., in the trachea \cite{varner2014cellular}) or reciprocal signalling with the surrounding mesenchyme (e.g., in the kidney \cite{mcmahon2016development}).

The third observation supporting our proposal is that whenever progenitor-cell-based morphogenesis evolved in our model, it had three processes that have been shown to elevate morphogenetic reproducibility \cite{cano2023origins}: (1) boundary-localised differentiation mediated by morphogen signalling, (2) the immobility and non-division of differentiated cells, and (3) differential adhesion between progenitor and differentiated cells. These processes have been shown to enhance morphogenetic reproducibility by smoothing the boundaries between domains of different cell types \cite{hagolani2019cell, cano2023origins, mizuno2024robust}, with smoother boundaries contributing to a more stable (i.e., less noisy) spatial distributions of cell types and, thus, more reproducible morphogenesis. Indeed, the boundaries between progenitor and differentiated cell domains in our evolved morphologies appear to be smooth (Figs.\,\ref{results}BDF, \ref{fungi}A and S1). We also found that these three processes enable differentiated cells to act as anchors that guide consistent progenitor-cell motion and differentiation, which again enhances morphogenetic reproducibility. These three processes are widespread at cell-type boundaries in extant developmental systems and are known to suppress noisy cell state distributions \cite{ inoue2001role, kicheva2014coordination, canela2011dynamics, mizuno2024robust}, implying they are generic processes of morphogenesis. Critically, we observed these three processes whenever progenitor-cell-based morphogenesis evolved, despite no direct selection for them. Therefore, progenitor-cell systems appear to be automatically equipped with these three processes, implying that progenitor-cell-based morphogenesis will be generally reproducible.



In summary, our results suggest that progenitor-cell systems, in addition to producing specialised cell types, can underpin multicellular morphogenesis and its reproducibility via cell differentiation. Expanding our focus on cell differentiation from solely cell-type specification to also encompassing morphogenesis has important implications for how we understand the generation and regeneration of tissue morphologies. The programming of tissue morphogenesis \textit{ex vivo}, such as organoids, in an accurate and reproducible way is a critical challenge in synthetic biology \cite{huch2017hope, toda2018programming, gjorevski2022tissue}. Our findings suggest that a better understanding of the role of cell differentiation in morphogenesis will enable us to more effectively manipulate the accuracy and reproducibility of programmed morphogenesis. Finally, our findings open the possibility that cell differentiation during the development of some animal tissues evolved for a morphogenetic purpose, with specialised cell types emerging as a later exaptation.

\section{Methods}

\subsection{Cellular Potts Model}
\label{cpm}

Our model extends a Cellular Potts Model (CPM) introduced by Hogeweg (2000) \cite{hogeweg2000evolving} by adapting an implementation of the Tissue Simulation Toolkit \cite{merks2005cell, daub2014cell}. The CPM dynamics are driven by pixel copying, where repeated random sampling of pixels on grid determines the location of these copies. For each chosen pixel, a random neighbouring pixel from its Moore neighbourhood is selected as the recipient of the pixel copy. Whether the pixel copy is accepted depends on its effect on the system's energy, denoted as $H$, represented by
\begin{equation}
    H = \sum_{i,j } {J_{ij}} + \sum_{i,m} {J_{im}} +  \lambda_V \sum_{\sigma} (\upsilon_{\sigma}-V_{\sigma})^2 + \lambda^{\sigma}_L \sum_{\sigma}  (l_{\sigma}-L_{\sigma})^2.
    \label{H}
\end{equation}
Here, $H$ encompasses the total surface energy accumulated from cell-cell adhesion ($J_{ij}$) and cell-medium adhesion ($J_{im}$), which are both determined dynamically as functions of protein concentrations (explained later). The index $i$ is of pixels at cell boundaries. The index $j$ is of pixels neighbouring $i$ that are occupied by a different cell from $i$. The index $m$ is of pixels neighbouring $i$ that are occupied by the medium. Every pixel on the grid has an associated value, $\sigma$, that represents either the cell that occupies that pixel ($\sigma \geq 1$) or the medium ($\sigma=0$). Each cell $\sigma$ ($\sigma \geq 1$) with current size $\upsilon_{\sigma}$ (in pixels) is constrained to size $V_{\sigma}$ with parameter $\lambda_V$ ($\lambda_V=0.5$ for all simulations). The longest axis $l_{\sigma}$ (in pixels) of each cell is constrained to length $L_{\sigma}$ with $\lambda^{\sigma}_L$. $V_{\sigma}$, $L_{\sigma}$ and $\lambda^{\sigma}_L$ are determined dynamically (explained later). 

If a pixel copy attempt increases $H$, it is accepted with probability $e^{- \frac{\Delta H}{T}}$, where $\Delta H$ is the change in $H$ made by a pixel copy, and $T$ is a temperature-like parameter that we arbitrarily fixed to $T=3$. Otherwise ($\Delta H\leq0$), the pixel copy attempt is always accepted. One developmental time step (DTS) is complete when the number of pixel copy attempts equals the number of pixels on the grid.

\subsection{Development}
\label{development}

A morphology starts as a single cell that undergoes six divisions equally spaced in time over the first 300 DTS. During the first 1,500 DTS (referred to as an equilibration phase), $J_{ij}=30$ and $J_{im}=40$ for all cells. After the equilibration phase, $J_{ij}$ and $J_{im}$ are determined by protein concentrations in the cells (described later). At the beginning of the equilibration phase, $V_{\sigma}$ is set to the initial size of each of the 64 cells and does not change throughout the equilibration phase. After the equilibration phase, cell growth and shrinkage can occur through modulation of the target cell size, $V_{\sigma}$. $V_{\sigma}$ increases when a cell is stretched, and decreases when a cell is squeezed, as follows. When $\upsilon_{\sigma} \geq V_{\sigma} + 3$, $V_{\sigma}$ is updated such that $V_{\sigma}=\upsilon_{\sigma}$. When $\upsilon_{\sigma} \leq V_{\sigma} - 16$, $V_{\sigma}$ is updated such that $V_{\sigma}=\upsilon_{\sigma}$. When the cell volume reaches or exceeds a threshold (i.e., $\upsilon_{\sigma} \geq 100$), the cell undergoes division perpendicular to its longest axis. One daughter cell retains the same index $\sigma$ as the mother cell, while the other daughter cell is assigned a new index $\sigma'$. After division, the target area of the daughter cells' are $V_{\sigma} = \upsilon_{\sigma}$ and $V_{\sigma'} = \upsilon_{\sigma'}$. Protein concentrations remain unchanged upon cell division. If $\upsilon_{\sigma}=0$, the cell dies.

\subsection{Gene regulatory network and morphogens}
\label{grn}

We model protein concentration dynamics using a system of ordinary differential equations that correspond to the reaction-only component of the Reinitz and Sharp model \cite{reinitz1995mechanism}. In each cell $\sigma$, there are $N_{genes}$ genes, indexed by $p$, that each encode a protein whose intracellular concentration is denoted by $x_p^{\sigma}$. With the exception of $$p \in \{1, \dots, N_{morph}\}$$ (which are morphogens, described later), the following equation determines the change in $x_p^{\sigma}$ over time $t$:
\begin{equation}
    \frac{dx_p^{\sigma}}{dt} = \frac{a}{1+e^{-\beta f_p(x)}} - bx_p^{\sigma} 
    \label{ode}
\end{equation}
with one unit of $t$ being one DTS. The first term on the right-hand side represents the increase in $x_p^{\sigma}$ due to gene expression, and is a sigmoidal function that depends on transcription factor regulation, $f_p(x)$ (described subsequently), with a maximum production rate $a$ and large $\beta$ ($=20$). The second term represents protein decay with rate $b$. We set $a=b$ for all $p$. This ensures that when a gene is activated ($f_p(x)>0$), the production term dominates and $x_p^{\sigma}\xrightarrow[]{}1$. Conversely, when a gene is inhibited ($f_p(x)<0$), the production term is negligible compared to the decay term and $x_p^{\sigma}\xrightarrow[]{}0$. Thus, the expression dynamics are a switch-like response where genes transition between ``on'' or ``off'', i.e., $x_p^{\sigma}$ equilibrates at 1 if the gene is constantly expressed or 0 if it is constantly not expressed. We assume that the timescale of gene expression is slower than the timescale of CPM dynamics. Thus, we set $a$ and $b$ to small values (specifically, $6.25\times10^{-3}$). Numerical integration of Equation \ref{ode} occurs with $\Delta t_1=40$ via the Euler method (we use a subscript because Equation \ref{EqDiffusion}, described subsequently, is numerically integrated at a different interval). We chose this value of $\Delta t_1$ instead of a smaller value to improve computational speed. The value of $\Delta t_1$ can be large because the rate parameters $a$ and $b$ are very small. The initial concentrations are $x_p^{\sigma}=1$ for transcription factors (except for the two that are asymmetrically distributed, which are either $x_p^{\sigma}=1$ or $x_p^{\sigma}=0$ depending on $\sigma$ at the four-cell stage), and $x_p^{\sigma}=0$ for all other proteins (described later). Integration of Equation \ref{ode} begins after the four-cell stage is reached (100 DTS), which is when the asymmetric distribution of transcription factors (TFs) occurs.

Function $f_p(x)$ in Equation \ref{ode} sums the regulatory effects of $N_{TF}$ TFs (including morphogens) as follows:
\begin{equation}
     f_p(x) = \left[ \sum^n_{p'} Z_{pp'}x_{p'}^{\sigma} \right] + \theta
     \label{summation}
\end{equation}
where $Z_{pp'}$ is the regulatory effect of TF encoded by gene $p'$ on the expression of gene $p$ ($Z_{pp'} \in \{0,\pm1,\pm2\}$). The TF encoded by $p'$ activates the expression of $p$ if $Z_{pp'}>0$, inhibits if $Z_{pp'}<0$ and has no effect if $Z_{pp'}=0$. The parameter $\theta$ sets the base level of gene expression. $\theta=-0.3$ for all simulations, so protein concentrations equilibrate at 0 when they are not regulated by any TFs.  






To model cell-cell signalling, the first $N_{morph}$ ($p=1, \dots, N_{morph}$) of the $N_{TF}$ TFs are morphogens. These morphogens diffuse between cells and into the surrounding medium. The concentration of morphogen $p$ on pixel $i$ on the grid, $x^i_p$, is determined by the following coupled ODE:
\begin{equation}
    \frac{d x_p^i}{d t} = D_p \nabla^2 x^i_p + \omega_p H \left( \sigma - 1 \right) x^{\sigma}_{p_{sig}} - \eta x^i_p
    \label{EqDiffusion}
\end{equation}
where $D$ is a diffusion constant, $\omega$ is a production rate, $\eta$ is a decay rate and $H(\sigma - 1)$ is the Heaviside step function that evaluates to one if $\sigma \geq 1$ at pixel $i$ (the pixel is occupied by a cell) and zero if $\sigma=0$ (the pixel is medium). The concentration $x^{\sigma}_{p_{sig}}$ represents a signalling protein that activates the expression of morphogen $p$. The concentration of signalling proteins is determined by Equation \ref{ode}. The operator $\nabla^2x^i_p$ is the Laplacian acting on $x^i_p$, which, in this context, is the difference between $x^i_p$ and the average $x^i_p$ in its von Neumann neighbourhood (i.e., four nearest-neighbouring pixels) divided by a space step $(dx)^2$, with $dx=1/250$ (250 is the length of the square grid in pixels). Numerical integration for Equation \ref{EqDiffusion} occurs with $\Delta t_2=1$. Equation \ref{EqDiffusion} is subject to the initial condition $x^i_p=0$ for all $i$ and $p$. The pixels at the boundary of the grid are subject to $x_p^i=0$ for all $t$ and $p$. The constants $D_p=8\times10^{-7}$, $\omega_p = 2.4\times10^{-3}$ and $\eta=2\times10^{-3}$ for all $p$ were used for the 126 simulations outlined in the main text. We chose these values for two reasons. The first is so that the maximum concentration is approximately 1 and thus similar to the concentrations of TFs per Equation \ref{ode}. The second is so that the characteristic diffusion length, $\sqrt{D_p / \eta dx}$, is similar to that of a paracrine morphogen, such as Wnts, which signal only to nearby cells \cite{ farin2016visualization, kerridge2016modular}. This characteristic diffusion length is five pixels, which is approximately the diameter of a cell. Since $x_p^i$ regulates genes in each cell $\sigma$, we average $x^i_p$ over all $i$ with the value $\sigma$ to obtain $x^{\sigma}_{p}$. However, we also ran 44 simulations where $D_p$ and $\omega_p$ mutate (see Fig.\,S9 and Methods \ref{evolution}), resulting in each morphogen having a unique characteristic diffusion length. During the equilibration period (DTS $< 1500$), the parameters $D_p, \omega_p, \eta$ and $\Delta t$ are multiplied by eight in order to speed up the time taken for morphogens to reach a steady state. The faster timescale does not have an effect on morphogenesis because there is no GRN-driven dynamics during the equilibration phase (Methods \ref{development}).

For the simulations presented in the main text, we fixed $N_{morph}=3$ and $N_{TF}=9$. In the Supplementary Information, where we explore variations in network size, the specific values for $N_{morph}$ and $N_{TF}$ are explicitly stated.


\subsection{Adhesion and membrane tension proteins}
\label{proteins}

Cells encode adhesion proteins constituting $N_{pairs}$ pairs of lock-and-key proteins that determine cell-cell adhesion and $N_{med}$ medium proteins that determine cell adhesion to the extracellular medium as well as cell fluidity (explained subsequently). While a simpler homotypic adhesion system, such as that mediated by E-cadherin \cite{lecuit2007cell}, could have been used, we chose this asymmetric lock-and-key system to provide a more flexible adhesion code to allow a larger array of evolvable interactions between cell states. Each lock protein has a complementary key protein to which it can bind. When two cells are in contact, the adhesion energy between them ($J_{ij}$ in Equation \ref{H}) decreases with the number of expressed pairs of compatible locks and keys (described subsequently). Each pair of lock and key reduces adhesion energy by the same amount. The adhesion energy to the extracellular medium ($J_{im}$ in Equation \ref{H}) decreases with the number of medium proteins expressed by a cell. Medium proteins have graded adhesion strengths (described subsequently). All $J_{ij}$ and $J_{im}$ variables used in our model are positive because cell disintegration is favoured when they are negative. When $J_{ij}$ and/or $J_{im}$ are small, the energy barriers to cell rearrangements are low and cells behave more fluid-like. When $J_{ij}$ and/or $J_{im}$ are high, the energy barriers to cell rearrangements are high and cells behave more solid-like.

Adhesion protein concentrations are booleanised to an ON or OFF state for adhesion energy calculations (i.e., ON if $x^{\sigma}_p > 0.5$ else OFF). Specifically, $J_{ij}$ between neighbouring pixels $i$ and $j$ that belong to different cells is:
\begin{equation}
    J_{ij} = J_{ij}^\text{max} -  2\sum_{k=1}^{N_{pairs}} \left[ \phi^{ij}_k
    + \phi^{ji}_k \right]
\end{equation}
where $\phi^{ij}_k=1$ if the $k$-th lock in the cell of pixel $i$ and the $k$-th key in the cell of pixel $j$ are both ON and otherwise $\phi^{ij}_k=0$. Similarly, $J_{im}$ between pixels $i$ and $m$ belonging to a cell and the medium, respectively, is:
\begin{equation}
    J_{im} = J_{im}^{\text{max}} - \sum_{k=1}^{N_{med}} k \psi_k 
\end{equation}
where $\psi_k=1$ if the $k$-th medium-adhesion protein is ON, and otherwise $\psi_k=0$. $J_{ij}^{\text{max}}=24$ and $J_{im}^{max}=21$ for all simulations. In Figure S14A-D, we show that the morphogenesis of evolved morphologies is not disrupted by changes to these parameters.

Cells encode $N_{tens}$ membrane tension proteins that make the cell shape less deformable by energetically constraining it to an elliptic shape. Consequently, cells become more solid-like when they express membrane-tension proteins. Cell shapes are defined as ellipses that have a major axis of length $l_{\sigma}$ constrained to a target length of $L_{\sigma}$ with $\lambda^{\sigma}_L$ (see Equation \ref{H}). $L_{\sigma}$ increases with the number of expressed membrane tension proteins. The major axis is defined as the longest dimension of the cell, irrespective of its orientation. Because the energetic constraint applies to this dynamically identified major axis without preference for a specific direction, there is no inherent polarity to the cell's contraction. The orientation and length of the major axis of a cell are re-evaluated after each pixel copy attempt involving that cell. Membrane tension proteins are first booleanised to an ON or OFF state by the same method used for adhesion proteins. For simulations in the main text, $N_{tens}=2$. When one length protein is ON in cell $\sigma$, $L_{\sigma}$ is set to $1.65\sqrt(\upsilon_{\sigma})$; When two are ON, $L_{\sigma}$ is set to one-third of $1.25\sqrt(\upsilon_{\sigma})$ (the numbers 1.65 and 1.25 are arbitrarily chosen). To implement the presence or absence of a length constraint depending on whether membrane tension proteins are expressed, we set $\lambda^{\sigma}_L=0.1$ if either membrane tension protein in cell $\sigma$ is ON, otherwise $\lambda^{\sigma}_L=0$. 

For the simulations presented in the main text, we fixed $N_{pairs}=5$, $N_{med}=5$ and $N_{tens}=2$. In the Supplementary Information, where we explore variations in network size, the specific values for $N_{pairs}$, $N_{med}$ and $N_{tens}$ are explicitly stated.

We calculated surface tensions for the 24 the morphologies that evolved progenitor-cell systems in the simulations presented in the main text. These calculations determined the surface tension for three distinct interfaces: (i) between each progenitor-cell type and the medium, (ii) between each differentiated-cell type and the medium, and (iii) between progenitor and differentiated cells. We determined surface tensions, $\gamma_{\tau \tau'}$ by the following formula:
\begin{equation}
    \gamma_{\tau \tau'} = J_{\tau \tau'} - \frac{J_{\tau \tau}+J_{\tau' \tau'}}{2T}
    \label{morpho-tension}
\end{equation}
where $J_{\tau \tau'}$ is the adhesion energy arising between cell types $\tau$ and $\tau'$, while $J_{\tau \tau}$ and $J_{\tau' \tau'}$ are adhesion energies arising from contact between two cells of the same type, $\tau$ and $\tau'$, respectively. For calculations involving the medium (i.e., when$\tau'=m$), the adhesion energy between units of the medium is zero ($J_{\tau \tau}=0$). The parameter $T$ is the temperature-like parameter governing the Metropolis algorithm, which scales all energy barriers in the CPM ($T=3$ for all simulations). For each surface tension calculation, we used the most frequently observed cell state to represent its corresponding cell type. The value of the surface tension between a cell type $\tau$ and the medium ($m$) predicts the morphology of cell clusters \cite{graner1992simulation}. If $\gamma_{\tau m}>0$, cell clusters will minimise their surface area and form circular aggregates. If $\gamma_{\tau m}<0$, cells maximise their contact with the medium and disperse from each other. As $\gamma_{\tau m} \xrightarrow[]{}0$, the interface with the medium becomes energetically neutral, and the shape of cell clusters is less constrained by adhesion.

\subsection{Evolution}
\label{evolution}

To simulate the evolution of morphogenesis, we established an initial population of 60 morphologies with each assigned a different GRN and developing on a separate CPM grid. Each GRN is specified by 234 $Z_{pp'}$ values representing the regulatory effects of all transcription factors, as described in Equation \ref{summation} (nine transcription factors regulate 26 genes including themselves). The $Z_{pp'}$ values of the GRNs assigned to the initial population are randomly generated according to the following probabilities: $P(Z_{pp'}=0)=0.54$, $P(Z_{pp'}=1)=0.18$, $P(Z_{pp'}=-1)=0.18$, $P(Z_{pp'}=2)=0.05$ and $P(Z_{pp'}=-2)=0.05$. These probability choices are arbitrary. The 15 morphologies with the highest shape complexity reproduce four times to populate the next generation (the definition of shape complexity is described in the next paragraph). Upon reproduction, there is a 50\% chance that one of the $Z_{pp'}$ values in the GRN mutates. The specific $Z_{pp'}$ that mutates is chosen at random with equal probability. The mutation alters the value of $Z_{pp'}$, independent of its current value, according to same probabilities used to generate the $Z_{pp'}$ assigned to the initial population. Gene duplication and deletion do not occur. 

In simulations where morphogen diffusivity mutates (Fig.\,S9), the values of the diffusion constant, $D_p$ (see Eq.\,\ref{EqDiffusion}) start as a random value taken from a normal distribution between $D_{min}=3\times10^{-8}$ and $D_{max}=8\times10^{-7}$. We also change the secretion constant, $\omega_p$ (see Eq.\,\ref{EqDiffusion}), to match changes in $D_p$. Specifically, the value of $\omega_p$ is equal to:
\begin{equation}
    \omega_p = \omega_{min}+\epsilon \frac{D_p-D_{min}}{D_{max}-D_{min}},
\end{equation}
where $\omega_{min}=2.4\times10^{-3}$ is the minimum value of $\omega_p$ and the second term makes $\omega_p$ increase with $D_p$, modulated by the constant $\epsilon=1.5\times10^{-3}$. The purpose of increasing $\omega_p$ with $D_p$ is to ensure that the maximum concentration of morphogen $x_p^i$ in Equation \ref{EqDiffusion} remains close to one. Only $D_p$, not $\omega_p$, affects the characteristic length of morphogen diffusion at equilibrium, as described in Methods \ref{grn}. Upon reproduction, there is a 25\% chance that one of the $D_p$ values mutates, with the choice of $p$ occurring
randomly with equal probability. The mutation alters the current value of $D_p$ by multiplying it by $e^{-\mu}$, where $\mu$ is a random number picked from a normal distribution of mean zero and standard deviation 0.25.

We defined shape complexity based on an algorithm that quantifies the complexity of two-dimensional shapes \cite{page2003shape}. Shape complexity is a summation of two measures: the deviation of morphology from a circle (denoted by $z_1$; Fig.\,S15A) and the degree of inward folds (denoted by $z_2$; Fig\,S15B), as described below.

The deviation of a morphology from a circle $z_1$ is defined by the following equation:
\begin{equation}
    z_1=\langle |r_c - r(\theta)| \rangle_{\theta}
\end{equation}
where $r_c$ is the hypothetical radius of the morphology if its mass were to be redistributed into a perfect circle, and $r(\theta)$ is the maximum distance from the centre of mass of the morphology to any pixel in the direction specified by angle $\theta$ (one pixel corresponds to one unit mass). The notation $\langle ... \rangle_{\theta}$ indicates an average taken over all angles $\theta$, where $\theta$ is discretised into 360 degrees for computation. 

The degree of inward folds ($z_2$) measures the sum of the sizes of regions of the medium surrounded by concave parts of a morphology. To identify these regions, we begin by drawing horizontal parallel lines across the CPM grid spaced one pixel apart, resulting in a total of 250 lines. Next, we located all segments along the lines that intersect the extracellular medium in between cells. The segments were discarded if they did not exceed a minimum length of 20 pixels to filter out inward folds due to stochastic cell boundary fluctuations. The above procedure was repeated by tilting the 250 parallel lines at 12 evenly-spaced angles across the range $[-\pi/2, \pi / 2]$. $z_2$ is defined as the square root of the total number of the located segments ($z_2$ is independent of the lengths of the retained segments).

We defined shape complexity as $2z_1 + 2.5z_2$. The weights of 2 and 2.5 were selected to ensure that the maximum value of either term is approximately 100 for shapes that are realistically achievable within this model, ensuring that neither term dominates the fitness criterion. In order to filter out noise, shape complexity is taken as an average of 10 evenly-spaced measurements over the last 1,000 DTS. If cells lose physical contact with other cells during development (see Fig.\,S14EF for examples), we assign the morphology a fitness of 0 as our quantification of shape complexity is not designed to handle multiple shapes on the same grid. 

We implemented an alternative fitness criterion to determine where progenitor-cell-based morphogenesis is evolutionary accessible (see Fig.\,S10), which has two stages. In the first stage, the fitness criterion is the displacement of a morphology's centre of mass, measured in pixels, from the start to the end of the 12,000 DTS, denoted $z_3$. Once the average fitness across all morphologies in a population exceeded 15 pixels, the fitness criterion transitioned to the second stage. In the second stage, the fitness criterion is ($2z_1+2.5z_2)/2 +z_3^{1.5}$. The weightings of each criterion were selected so that the maximum value of each is approximately 100 for shapes that are realistically achievable within this model, ensuring that neither term dominates the fitness criterion.

To determine whether an evolutionary simulation succeeded or failed, we used fitness thresholds. The thresholds were applied to the morphology that recorded the most complex morphology in the final generation of the simulation. In the simulations selecting only for shape complexity, we applied a threshold of 70 after averaging the shape complexity over 60 developmental replicates in order to account for variance in the complexity score. However, this threshold does not affect our key findings, as there are still highly reproducible morphologies with progenitor-cell systems below the threshold (See Fig.\,S2AB). In the simulations selecting for both shifting centre of mass and shape complexity, the fitness threshold we used to determine whether an evolutionary simulation succeeded was whether the second stage was reached.





\subsection{Cell state and state space}
\label{statespace}

The cell state is an $n$-dimensional boolean vector, where $n$ is the total number of adhesion and membrane tension proteins (although the results do not change when TFs are included in the vector as well). Each element in this vector is the concentration of each protein booleanised to either ON or OFF (as described previously for lock, key and tension proteins). Each boolean vector is assigned a single colour on the CPM grid. These colours are chosen arbitrarily, and each morphology has its own distinct colour set. While this approach means that the same cell state might appear in different colours across morphologies, it is uncommon to encounter identical cell states in different evolved morphologies. The cell state of each cell is determined after every numerical integration step of Equation \ref{ode}. A change in a cell's boolean vector corresponds to a cell state transition.  

To generate the cell state space, we recorded all cell state transitions for all cells after 6,000 DTS from the beginning of development. Although it makes no qualitative difference when the recording of transitions begins, starting at 6,000 reduces the appearance of ``transient'' cell states \cite{wuensche2004basins}, such as the cell state corresponding to the initial conditions of cell proteins, which makes it easier to visualise the cell state space. Cell state transitions are recorded from 10 developmental replicates in order to reduce the effect of noise on cell state space creation. The state space is the directed graph, where the nodes are the cell states and the edges are the transitions. Figure S5IJ shows two examples of these directed graphs. To simplify the graph, we prune rare transitions (those that occur less than five times across all cells per developmental replicate) and rare cell states. To identify rare cell states, we first count the number of cells in each state after each integration step of Equation \ref{ode} to obtain a frequency distribution of cell states. Rare cell states are those that have a frequency of less than 1\% from the 10 developmental replicates. However, even without any pruning of cell state transitions and cell states, the cell state space of 22 out of 24 morphologies designated as having progenitor-cell systems still exhibit irreversible differentiation (Fig.\,S14G-J shows one that does not follow this rule). 

We used a depth-first search algorithm to identify a graph's strongly connected components (SCCs). Irreversible differentiation occurs when there is weak connectivity between SCCs (i.e., a path exists from SCC $u$ to SCC $v$, but not $v$ to $u$). SCCs that do not have incoming paths from any other SCCs (source components) are designated as progenitor-cell types. SCCs that have no outgoing paths to any other SCCs are designated as differentiated-cell types.



\subsection{Reproducibility score}
\label{measurements}
We measured the morphogenetic reproducibility of morphologies in a rotation-, reflection- and translation-invariant manner, as follows. We first replayed the development of a morphology 60 times with different random seeds. We then performed pairwise comparisons of all developed morphologies ($60\times59/2$ comparisons). For each pair of morphologies, we computed morphological similarity scores between the two CPM grids on which morphologies developed (denoted by $A$ and $B$). We computed the morphological similarity scores (denoted $\text{Jac}$) over many rotations, reflections and translations of grid $B$ relative to a fixed grid $A$ to find the maximum possible similarity between them (denoted $\text{Jac}_\text{max}$). 

To calculate $\text{Jac}$, grids $A$ and $B$ are transformed from Cartesian to polar coordinates, as follows. The polar coordinate $(r,\theta)$ is discretised into $250\times360$ pixels. Each pixel in $(r,\theta)$ is mapped to the pixel closest to $(r\cos\theta, r\sin\theta)$ in $A$ or $B$, where $r$ is the distance from the midpoint of grid $A$ or one of 25 equally-spaced locations in grid $B$ (thus, multiple pixels in the polar coordinate can be mapped to the same pixel in the Cartesian coordinate). Let $A_{\theta}^r$ and $B_{\theta}^r$ be one if the mapped pixel in $A$ or $B$ belongs to a biological cell and 0 otherwise. We then calculated the Jaccard index, $\text{Jac}(A_{\theta}^r,B_{\theta}^r)$, which measures the similarity between $A_{\theta}^r$ and $B_{\theta}^r)$, as follows:
\begin{equation}
    \text{Jac}(A_{\theta}^r,B_{\theta}^r) = \frac{ A_{\theta}^r \cap B_{\theta}^r}{ A_{\theta}^r \cup B_{\theta}^r}
\end{equation}
where $A_{\theta}^r \cap B_{\theta}^r=\sum^{359^\circ}_{\theta=0^{\circ}} \sum_{r=0}^{249} r \delta(A_{\theta}^r,1) \delta( B_{\theta}^r,1)  $ is the number of pixels in the polar coordinate with a value of one (i.e., the pixel is occupied by a cell) on both $A_{\theta}^r$ and $B_{\theta}^r$, where $\delta$ is the Kronecker delta. The multiplication by $r$ accounts for the fact that the length of an arc drawn by an increment in $\theta$ increases as $r$ increases. Similarly, $ A_{\theta}^r \cup B_{\theta}^r=\sum^{359^\circ}_{\theta=0^{\circ}} \sum_{r=0}^{249}r [\delta(A_{\theta}^r,1) + \delta( B_{\theta}^r,1)   -\delta (A_{\theta}^r, B_{\theta}^r)]$ is the number of pixels in the polar coordinate in which $A_{\theta}^r=1$ or $B_{\theta}^r=1$ (or both). Thus, when $A_{\theta}^r$ and  $B_{\theta}^r$ are exactly the same, $\text{Jac}(A_{\theta}^r,B_{\theta}^r)=1$. Next, we shift all values of $B_{\theta}^r$ to $B_{\theta+1 \pmod{360}}^r$, corresponding to a one-degree rotation in Cartesian coordinates, and compute $\text{Jac}(A_{\theta}^r,B_{\theta}^r)$ again. We repeat these one-degree rotations 360 times, computing $\text{Jac}(A_{\theta}^r,B_{\theta}^r)$ for each. Next, we invert all values of $B_{\theta}^r$ to $B_{180 - \theta \pmod{360}}^r$, which corresponds to a reflection of grid $B$, and repeat the 360 one-degree rotations again, computing $\text{Jac}(A_{\theta}^r,B_{\theta}^r)$ for each. Furthermore, we repeated these 720 computations for 25 equally-spaced translations of grid $B$ (as mentioned previously), achieved by shifting the location on grid $B$ chosen to be $r=0$ for the polar coordinate. Translations occur in steps of five pixels at a time to make up a five-by-five square. The maximum $\text{Jac}(A_{\theta}^r,B_{\theta}^r)$ recorded across all rotations, reflections and translations for each pairwise comparison is the maximum possible similarity, denoted $\text{Jac}_\text{max}$. The reproducibility score for a morphology is the average $\text{Jac}_\text{max}$ across all $60\times59/2$ pairwise comparisons.




\subsection{Momentum and anisotropy} 
\label{Momentum}
We defined the momentum, $p_{\sigma}(t)$, of cell $\sigma$ at time $t$ (where $t$ is in DTS) as the distance travelled by the cell's centre of mass between $t-t_w$ and $t$ multiplied by the cell's mass at $t-t_w/2$, with unit mass represented by one pixel:
\begin{equation}
    p_{\sigma}(t) = m_{\sigma}\left(t-\frac{t_w}{2}\right) \frac{s_{\sigma}(t) - s_{\sigma}(t-t_w)}{t_w}
    \label{momentEq}
\end{equation}
 where $s_{\sigma}(t)$ is the cell's centre of mass, $m_{\sigma}(t$) is the cell's mass at $t$, and $t_w$ is the waiting time. We set $t_w$ to 500 in order to average out the influence of stochastic membrane fluctuations and cell divisions on the centre of mass of the cell, thereby capturing the true mobility of cells. If a cell divides during the waiting time and the $\sigma$ values of the parent and daughter cells differ, Equation \ref{momentEq} is modified by subtracting the position of the parent cell's $\sigma$ centre of mass ($s_{\sigma}(t-t_w)$) from the daughter cell's $\sigma'$ centre of mass ($s_{\sigma'}(t)$.

In order to separate cell momentum by SCC, we first connected each recording of cell momentum $p_{\sigma}(t)$ to the state of the cell at $t-\frac{t_w}{2}$. We then assigned each momentum recording into an SCC depending on that cell state. To create the polar plots of cell momentum magnitude (Fig.\,\ref{fungi}DE), we categorised each momentum measurement assigned to an SCC into one of 36 bins based on its direction. Each bin encompasses momentum measurements within an angular width of 10$^\circ$. To measure anisotropic motion (Fig.\,\ref{fungi}FGH), we calculated the variance across these 36 bins. To account for total momentum, we divided the variance by the mean momentum across the 36 bins, equal to the dispersion index. The anisotropy value for an SCC is the average of the dispersion indices across five developmental replicates.



\bibliography{bibliography}
\bibliographystyle{unsrt}

\section*{Contributions}
D.K.D. conceived and designed the study, implemented the model, performed the analyses, and drafted the manuscript. N.T. contributed to the study design, provided feedback on the results and implications of the study, and commented on the manuscript at all stages. A.R.D.G. provided feedback on the results and implications of the study, and commented on the manuscript at all stages.

\section*{Competing interests}

The authors declare no competing interests.

\subsection*{Materials \& Correspondence}
\begin{itemize}
    \item All original code to generate the data in this study is publicly available, and can be found at \url{https://github.com/DominicDevlin/Stem-cell-differentiation-underpins-reproducible-morphogenesis}. Our code was adapted from the Tissue Simulation Toolkit \cite{daub2015cell}.
    \item The genomes and associated data of all evolved morphologies are also found in the above repository.
    \item Any additional information required to re-analyse the data reported in this paper is available from the lead contact upon request (dominicdevlin@g.ecc.u-tokyo.ac.jp)
\end{itemize}

\end{document}


\maketitle



\renewcommand{\abstractname}{} 
\begin{abstract}
\vspace{2cm}
\hspace{-0.5cm}\textbf{The Supplementary information includes:}
\begin{itemize}
    \item Text S1 and S2
    \item Legends for Movies S1 to S6
    \item Figures S1 to S15
    \item References for the supplementary text
\end{itemize}

\medskip
\hspace{-0.5cm}\textbf{Other supplementary materials for this manuscript include the following:}
\begin{itemize}
    \item Movies S1 to S6, found at \url{https://dominicdevlin.github.io/Cell-differentiation-Supplementary-Videos/}
\end{itemize}
\end{abstract}

\clearpage

\section{Morphologies with divisions of labour have higher reproducibility when accounting for differences in shape complexity}
\label{SuppReproducibility}

This section aims to determine whether the difference in reproducibility scores between evolved morphologies with and without morphogenetic divisions of labour can be attributed to variations in their shape complexity. Morphogenetic divisions of labour are defined by the presence of multiple SCCs, with moving and stationary cell states separated into distinct SCCs (see main text). To demonstrate the relationship between shape complexity and reproducibility in our model, imagine the simplest morphological shape: one that remains circular over development. The reproducibility of this morphology will be trivially high because its circular morphology will not change in each developmental replicate. In contrast, morphogenesis requires extensive cell motion. Extensive cell motion is prone to noise in cell motion and geometry, which means that morphogenesis tends to become more susceptible to noise with increasing complexity of these cell movements. This relationship between complexity and reproducibility has also been demonstrated in a previous study \cite{hagolani2019cell}.

To answer whether differences in reproducibility scores between evolved morphologies with and without divisions of labour can be attributed to variations in their shape complexity, we first compared the average shape complexity between the two groups using the technique outlined in Methods 4.5, averaged over 60 developmental replicates. We found that those with divisions of labour ($\mu=94.1$, $\sigma=13.3$) exhibit lower shape complexity on average than those without ($\mu=116.5$, $\sigma=18.0$), and that this difference was statistically significant ($p=10^{-4}$, two-tailed t-test). Although our quantification of shape complexity is arbitrary, this result suggests that the difference in reproducibility scores may in part be explained by differences in shape complexity. 

We conducted two further analyses to more rigorously determine whether this difference in shape complexity was responsible for the difference in reproducibility scores. For the first analysis, we examined whether ``within-group similarity'' differed significantly between morphologies with and without morphogenetic divisions of labour. Within-group similarity measures how similar the morphologies are within a group (with or without division of labour). To quantify within-group similarity, we computed the morphological similarity of one developmental replicate of each morphology within a group at 12,000 developmental time steps (DTS) to all others within the same group (determined by the maximum Jaccard index; see Methods 4.7). Suppose the morphologies within a group have low complexity. In this case, they will deviate little from the initial circular shape, resulting in higher morphological similarity and, thus, higher within-group similarity scores. Conversely, more complex morphologies will result in morphological dissimilarity and, thus, lower within-group similarity scores. The advantage of using within-group similarity to determine the link between reproducibility and complexity is that it can be compared to reproducibility scores, as it uses the same technique, while avoiding our arbitrary quantification of shape complexity. The results show that morphologies with divisions of labour had a marginally higher within-group similarity (mean=45.5\%) than morphologies without (mean=43.9\%, $p=0.036$ two-tailed t-test; Fig.\,\ref{tradeoff}D). Despite this marginally higher similarity, there was a significant overlap in the interquartile ranges between the two groups. Moreover, the difference in means of within-group similarity scores is much smaller than the difference in reproducibility scores, which were, on average, 72.1\% for morphologies with divisions of labour versus 52.0\% for morphologies without ($p<10^{-12}$, two-tailed t-test). This result indicates that the difference in shape complexity between the groups does not fully account for their reproducibility differences.

For the second analysis, we conducted a regression to examine whether the relationship between shape complexity and reproducibility differs between those with and without morphogenetic divisions of labour. The regression model we used is:
\begin{equation}
R_i = \beta_0 + \beta_1 \cdot {M}_i + \beta_2 \cdot ({M}_i \times \text{DOL}_{i}) + \epsilon_i
\end{equation}
where $M_i$ and $R_i$ are the shape complexity and reproducibility of morphology $i$, respectively. The term ``DOL$_i$'' evaluates to one if morphology $i$ has a division of labour; otherwise, zero. We found that reproducibility declines more rapidly as complexity increases for those without divisions of labour (95\% confidence interval on $\beta_1$ is $[-3.0\times10^{-3}, -1.3\times10^{-3}]$, $n=65$), compared to those with them (95\% confidence interval on $\beta_1 + \beta_2$ is $[-1.2\times10^{-3}, 1.9\times10^{-5}]$, $n=24$). Moreover, when morphologies are matched for shape complexity, the graph shows no obvious overlap between the two groups (Fig.\,\ref{tradeoff}D). This result indicates that the elevated reproducibility observed in morphologies with divisions of labour is not because of their lower shape complexity than those without such systems.



The above result that reproducibility declines less with complexity in morphologies that have divisions of labour compared to those without indicates that division of labour allows morphogenesis to bypass a reproducibility-complexity trade-off. Ascertaining the existence of this trade-off from our regression analysis is difficult because the sample sizes we used are small (only 90 evolved morphologies). Therefore, we increased our sample size by taking samples of many morphologies from each simulation. We measured the reproducibility and average shape complexity of the morphologies with the highest fitness in each population at 100-generation intervals throughout each simulation. We analysed 36 simulations: the 18 that evolved morphogenetic divisions of labour and 18 that did not (we chose the latter based on having similar endpoint shape complexity to the former). We performed a linear regression of reproducibility against shape complexity for each simulation separately. The results show that reproducibility declined with complexity in every simulation (Fig.\,\ref{tradeoff}F), indicating that morphologies become less reproducible as they became more complex. However, the slopes appear steeper in simulations where poorly reproducible morphologies evolved (Fig.\,\ref{tradeoff}F). To quantify this, we bootstrapped the coefficients of the linear regression slopes to obtain a confidence interval of this coefficient for the 18 simulations where morphogenetic division of labour evolved and the 18 where they didn't. We found that this coefficient was much smaller in simulations where morphogenetic division of labour evolved ($95\%$ CI of slope is $-0.0017$ to $-0.0011$, Fig.\,\ref{tradeoff}FG, blue lines) compared to simulations where it did not ($95\%$ CI of slope is $-0.0035$ to $-0.0029$, Fig.\,\ref{tradeoff}FG, orange lines), indicating that reproducibility declines more rapidly as complexity increases without morphogenetic division of labour. Moreover, the ability of shape complexity to predict a morphologies' reproducibility was much weaker in those that evolved morphogenetic division of labour (average $R^2=0.59$ across the 18 simulations), compared to those that did not (average $R^2=0.91$ across the 18 simulations). These findings support the hypothesis that morphogenesis via a division of labour bypasses a trade-off between shape complexity and reproducibility. Moreover, the fact that differences in reproducibility persist through evolution suggests that simulations where divisions of labour evolved are on different evolutionary trajectories than those where they did not (these trajectories are likely determined by the initial conditions or early in the evolutionary simulations).

\clearpage

\section{Progenitor-cell systems can evolve from most initial gene regulatory networks}
\label{consistent}

Our results show that morphogenetic divisions of labour with irreversible differentiation, i.e., progenitor-cell systems, evolved only in a minority of simulations in the main text (24 out of the 90 morphologies with complex shapes). We asked whether this frequency is because progenitor-cell systems evolve from only a restricted portion of genotype space, and thus determined by the initial conditions, or whether the frequency depends on the selection pressure. To address this, we tested an alternative selection criterion that favoured not only the shape complexity but also the directional motion of morphologies. This additional selection for directional motion is expected to favour progenitor-cell systems because directional motion is a property of morphogenesis with progenitor-cell systems (Fig.\,5F in the main text), although it does not directly select for progenitor-cell systems or reproducibility. We quantified directional motion by determining how much a morphologies' centre of mass (measured in pixels) shifts over the 12,000 DTS. The selection pressure we used was an additive combination of this directional motion and the original quantification of shape complexity (Methods 4.5). We ran 35 simulations lasting at least $2.5\times10^3$ evolutionary generations, of which 31 evolved morphologies surpassed our arbitrary threshold of shape complexity (Methods 4.5). We found the great majority of these morphologies (29 out of 31) had evolved progenitor-cell systems (Fig.\,\ref{asym}ABCDG). Of these 29 morphologies, all but one displayed highly reproducible morphogenesis (Fig.\,\ref{asym}D). In contrast, the two morphologies that did not evolve progenitor-cell systems displayed poorly reproducible morphogenesis (Fig.\,\ref{asym}DH). In simulations where we only selected for directional motion and not shape complexity, progenitor-cell systems evolved in seven out of 25 simulations (Fig.\,\ref{asym}DEF). These results show that progenitor-cell systems can evolve from most initial gene regulatory networks.

\section*{Supplementary Movie Legends}
\textbf{Movie S1: Morphology-1 development for the 12,000 DTS.}

\textbf{Movie S2: Morphology-2 development for the 12,000 DTS.}

\textbf{Movie S3: Morphology-6 development for the 12,000 DTS.}

\textbf{Movie S4: Morphology-11 development for the 12,000 DTS.}

\textbf{Movie S5: Development of a morphology that creates its shape via cell death for the 12,000 DTS.}

\textbf{Movie S6: Development of a morphology that branches for the 12,000 DTS.}

\section*{Supplementary Figures}

\begin{figure}
    \centerfloat
    \includegraphics{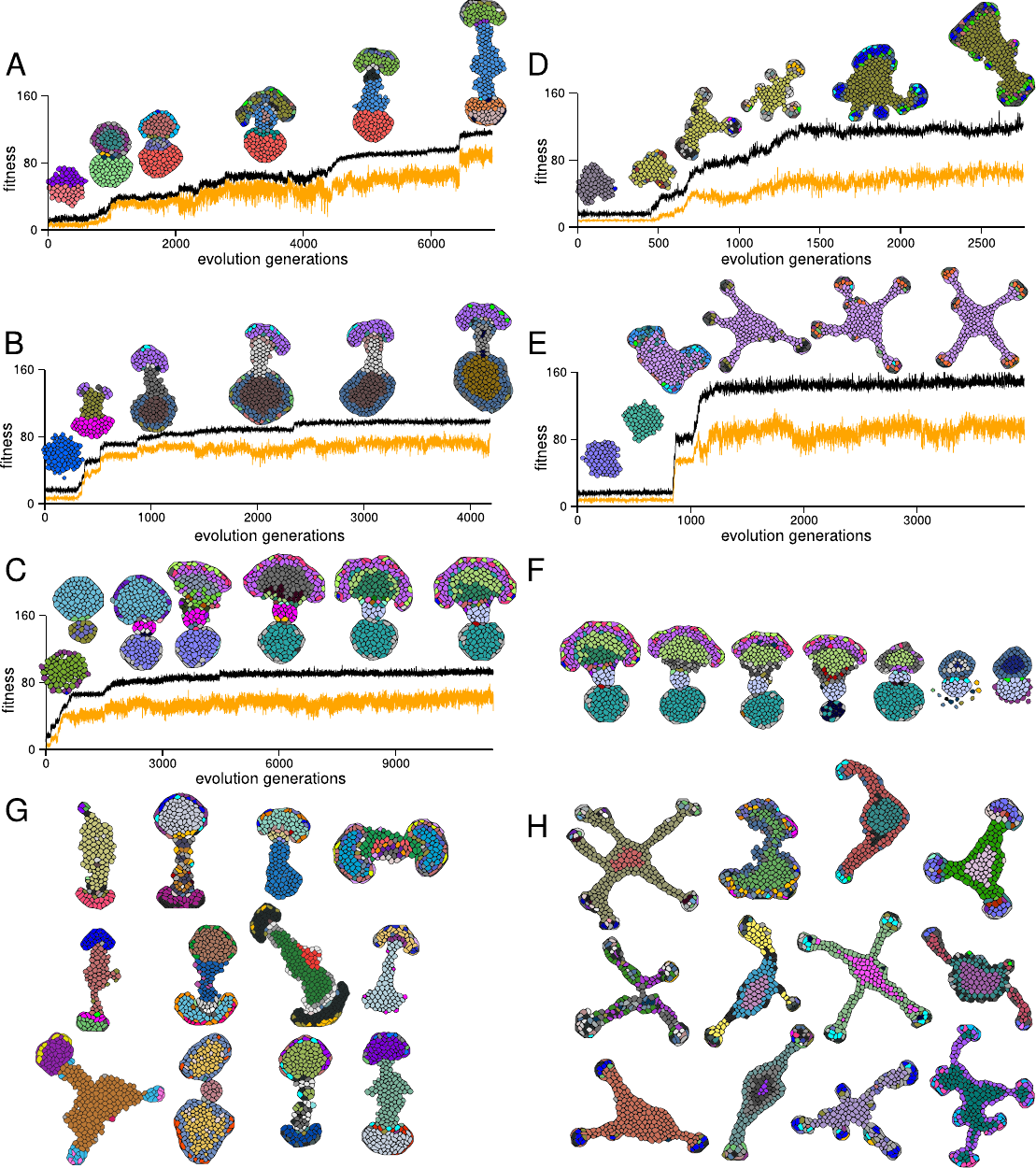}
\end{figure}
\clearpage
\captionof{figure}{\textbf{Evolution and evolved morphologies. A-C} Three plots of fitness evolution, each from a separate simulation where morphologies with high reproducibility evolved. The black lines show the maximum recorded fitness at each generation. The orange lines show the average fitness of each generation. Above each plot are six or seven morphologies at the end of their development (12,000 DTS). Each morphology is taken from a separate generation and is the most morphologically complex in its generation, shown approximately above the generation it belongs to. \textbf{DE} Two plots of fitness evolution, each from a separate simulation where morphologies with poor reproducibility evolved, mirroring the plots shown in (A-C). The evolutionary trajectories shown in (ABCDE) indicate that fitness reaches a plateau before $2.5\times10^3$ generations in 5 out of 6 simulations (we ran all simulations for at least this number of generations). Plateaus appear to persist for a long time, as exemplified by (C), which we ran for 11,500 generations. We did not run simulations for $>$10,000 generations due to computational limitations (10,000 generations takes approximately one week with 60 CPUs). The similarity of the morphologies above each plot indicates that the phenotype becomes conserved through evolution. This fixation implies that the evolved morphologies accurately represent the evolutionary outcome of a simulation. \textbf{F} Morphological variation in a population due to mutation. Each morphology is from the final generation from the simulation shown in (C). Each morphology has a unique GRN topology due to mutations. \textbf{G} Zoo of highly reproducible and \textbf{H} poorly reproducible morphologies, each shown at the end of their development (12,000 DTS). Each morphology is a different evolved morphology not shown in the main text.}
\label{stem-evolution}

\clearpage

\begin{figure}
    \centerfloat
    \includegraphics{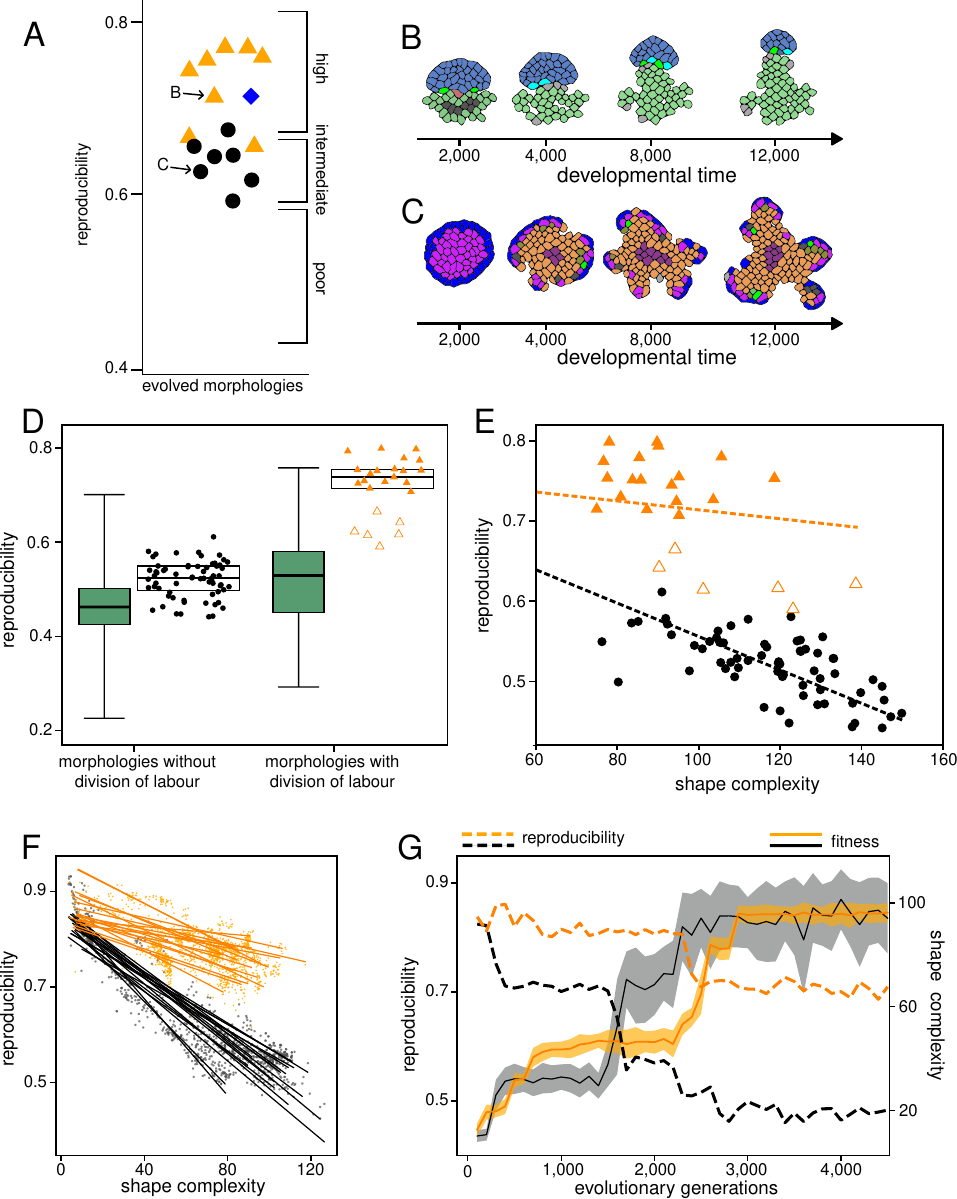}
\end{figure}
\clearpage
\captionof{figure}{\textbf{The trade-off between shape complexity and reproducibility. A} Reproducibility scores of 16 of the 36 ``failed evolved morphologies'' that did not reach the threshold for shape complexity. We chose these 16 because their shape complexity is above a score of 40 but below 70 (70 is the threshold we used to determine if a morphology has a complex morphology in the original 126 simulations). The remaining 19 morphologies with scores below 40 were excluded because they have circular morphologies. Black circles are morphologies that have state spaces with a single SCC, of which there are seven. The morphology indicated by the blue diamond has a state space with multiple SCCs with no unidirectional transitions. It has a similar morphogenesis to the morphology shown in Fig.\,\ref{transitions}D. Orange triangles are morphologies with progenitor-cell systems, of which there are eight. Categories of ``high'', ``intermediate'' and ``poor'' are copied from Fig.\,3C in the main text. The arrows point to the morphologies shown in (B) and (C). The eight morphologies with progenitor-cell systems were significantly more reproducible than those without ($p=0.007$, two-sample t-test). \textbf{D} Within-group reproducibility scores of the 65 evolved morphologies without progenitor-cell systems (left green box plot, $n=2211$) and the 24 with progenitor-cell systems (right green box plot, $n=276$). Boxes show medians and interquartile ranges (IQR). Whiskers show the range. Filled Orange triangles are reproducibility scores of morphologies with progenitor-cell systems that are highly reproducible; unfilled orange triangles are morphologies with progenitor-cell systems that are not highly reproducible. The black circles are morphologies without progenitor-cell systems. The unfilled box plot shows the median and interquartile range of reproducibility scores for morphologies with (right) and without (left) progenitor-cell systems. \textbf{E} Scatter plot of reproducibility scores against shape complexity for the 90 evolved morphologies. Data points are coloured and shaped as in (D). The black and orange dashed lines are the regression for poorly reproducible and highly reproducible morphologies described in Text S1. \textbf{F} Scatter plot of reproducibility scores against shape complexity scores. Each data point is the most complex morphologies in a population from a generation, taken at intervals of 100 generations. The orange data points are morphologies from simulations where highly reproducible morphologies with progenitor-cell systems evolved (18 simulations, 1508 data points), whereas the black data points are from simulations where poorly reproducible morphologies without progenitor-cell systems evolved (18 simulations, 1180 data points). Each line is the linear regression performed on all data points from the same simulation, coloured in the same way as the data points. Performing a single linear regression across all orange data points gives a 95\% confidence interval on the slope of $-0.0011$ to $-0.0099$ ($R^2=0.32$). Performing a single linear regression across all black data points gives a $95\%$ CI on the slope of $-0.0032$ to $-0.0031$ ($R^2=0.86$). \textbf{G} Evolution of reproducibility (dotted lines) and fitness (solid lines) in a simulation where a progenitor-cell system evolved (orange) and a simulation where one did not (black). The data used to generate the lines is the shape complexity and reproducibility of the highest-fitness morphology in populations at intervals of 100 generations for each simulation. A progenitor-cell system is first observed in the simulation in orange at generation 500.}
\label{tradeoff}


\clearpage

\begin{figure}[t]
    \centerfloat
    \includegraphics{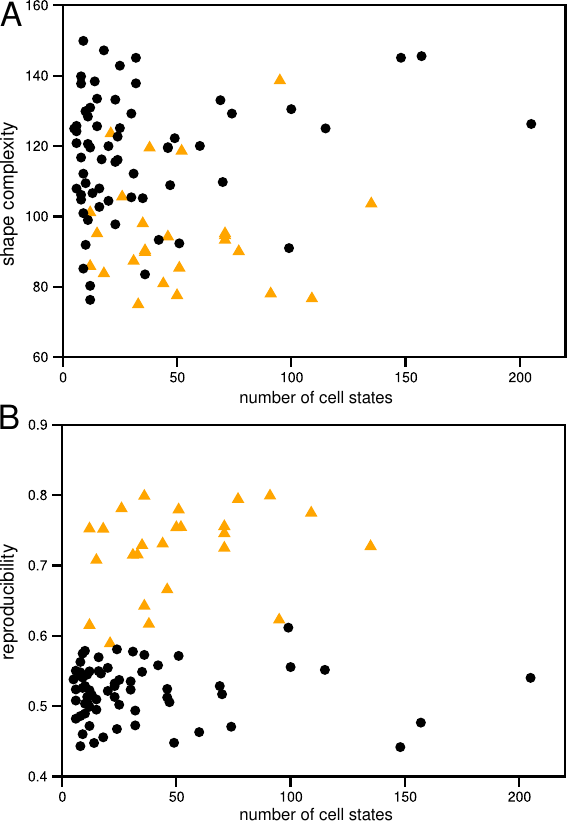}
    \caption{\textbf{Number of developmental cell states does not predict shape complexity or morphogenetic reproducibility. A} Shape complexity of each evolved morphology as a function of number of cell states. \textbf{B} Reproducibility score of each evolved morphology as a function of number of cell states. A cell state is counted towards the total if it is observed at any point in the development of a morphology over 10 developmental replicates, excluding the four possible initial states. Orange triangles are those classified as having high or intermediate reproducibility in the main text (Fig.\,3A). Black circles are those classified as having poor reproducibility in the main text (Fig.\,3A). For morphologies classified as highly reproducible (Fig.\,3A), a regression of shape complexity against number of cell states gives $R^2=0.02$, with a non-significant p-value on the slope ($p=0.56$). A regression of reproducibility against number of cell states gives $R^2=0.16$, with a non-significant p-value on the slope ($p=0.07$). For those classified as poorly reproducible morphology, a regression of shape complexity against number of cell states gives $R^2=0.04$, with a non-significant p-value on the slope ($p=0.12$). A regression of reproducibility against number of cell states gives $R^2=0.00$, with a non-significant p-value on the slope ($p=0.84$).}
\end{figure}
\clearpage

\begin{figure}
    \centerfloat
    \includegraphics{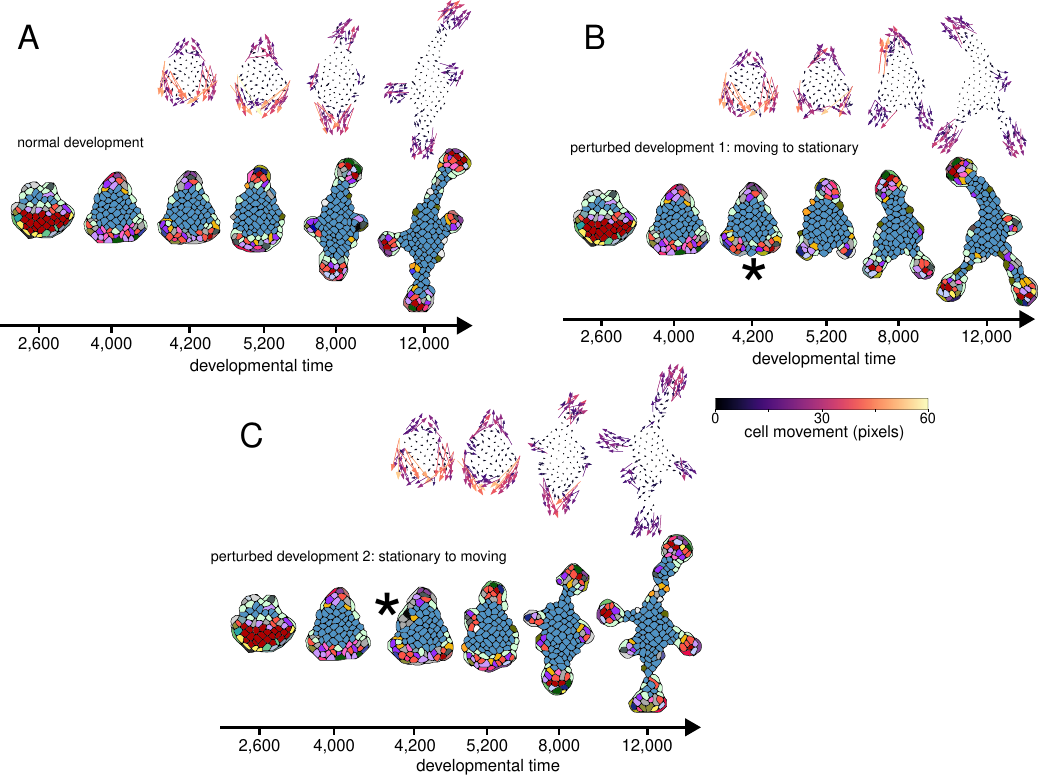}   
    \caption{\textbf{The causes of poor morphogenetic reproducibility. A} Developmental replicate of morphology-1 from the main text depicted at six DTS. Vector plots show the displacement of the centre of mass of each cell during 2,000 DTS at each respective time point, with colours indicating magnitude (the lighter, the larger). \textbf{B} A perturbed development of morphology-1 using the same random seed as (A) to show that moving-to-stationary transitions cause bifurcations. Development was perturbed by artificially changing the protein expression profiles of three moving cells at the bottom of the morphology to a stationary cell state (blue cells, indicated by the asterisk). We performed this perturbation after 4,100 DTS. This perturbation resulted in a bifurcation of collective cell motion, as indicated by the vector plots and the morphology. \textbf{C} A perturbed development of morphology-1 using the same random seed as (A) to show that stationary-to-moving transitions cause protrusions. Development was perturbed by artificially changing the protein expression profiles of three stationary cells at the left flank of the morphology to moving states (grey cells, indicated by the asterisk). We performed this perturbation after 4,100 DTS. This perturbation resulted in a protrusion, as indicated by the vector plots.}
    \label{transitions}
\end{figure}
\clearpage

\begin{figure}
    \centerfloat
    \includegraphics{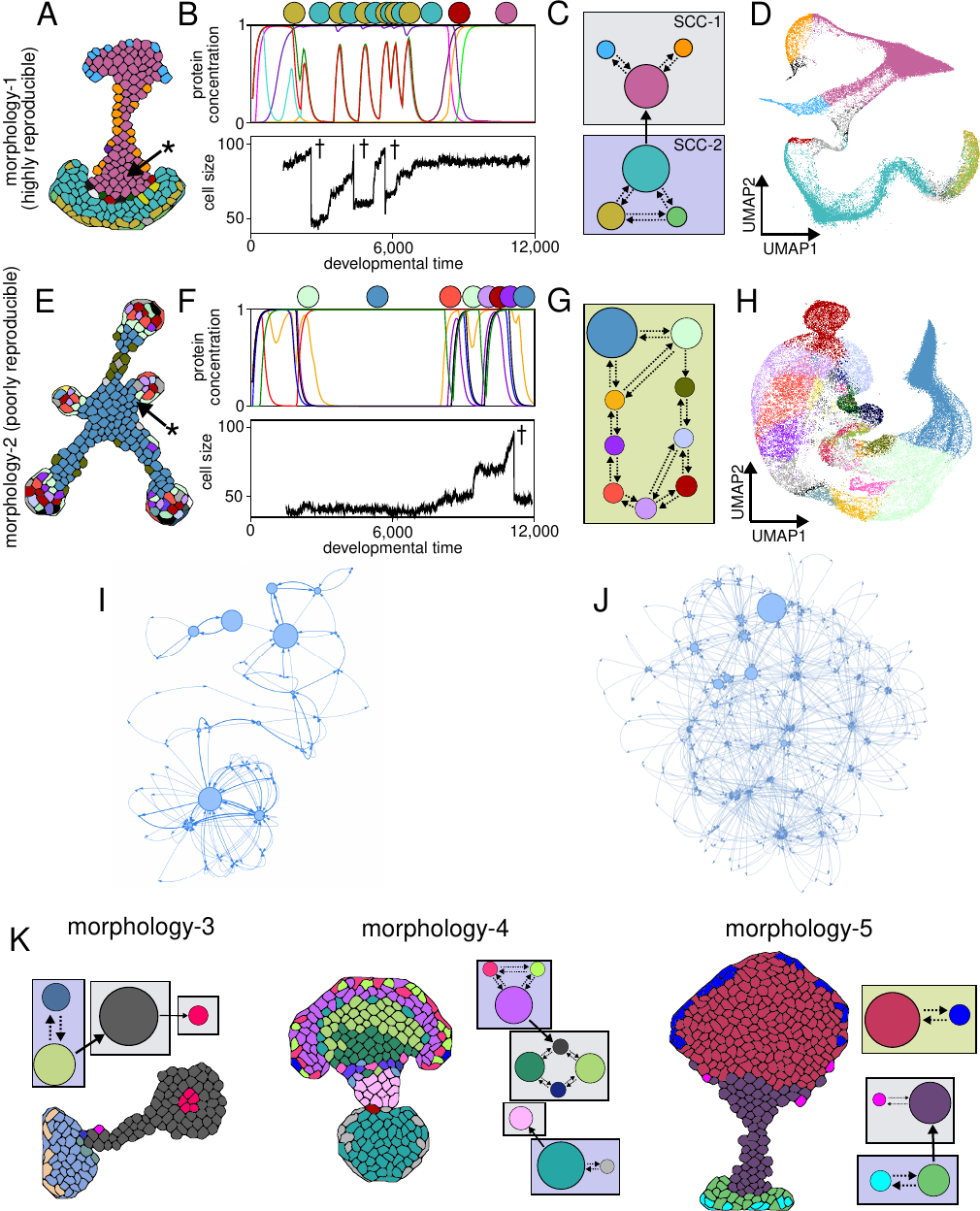}    
\end{figure}
\clearpage
\caption{\textbf{Highly and poorly reproducible morphologies have different cell-state transition dynamics.} Panels A through D correspond to morphology-1 (highly reproducible), and panels E through H to morphology-2 (poorly reproducible). \textbf{AE} Morphologies of morphology-1 and morphology-2 after the 12,000 DTS. \textbf{BF} Above the graphs are coloured circles depicting cell-state transitions for one cell, indicated by the asterisks in panels A and E, over the 12,000 DTS (some cell-state transitions are excluded for visibility). The top graphs show the concentrations of eight cell adhesion proteins over the 12,000 DTS. The bottom graphs show cell size over the 12,000 DTS. Daggers ($\dag$) indicate drops in size caused by cell division. \textbf{CG} Simplified cell state spaces consisting of cell states (nodes) and cell-state transitions (arrows). Node sizes depict cell state frequency. Cell states are partitioned into SCCs (coloured boxes). \textbf{DH} Visualisations of cell states (colours) mapped onto cell protein expression profiles (data points) that have undergone dimension reduction by UMAP. The data consists of 85,825 (D) and 115,625 (H) points, each collected from every cell at intervals of 40 DTS. \textbf{I} The state space shown in (C) without any pruning of cell states and cell state transitions. Many SCCs in the unpruned cell state space are ``transitory''. These transitory SCCs consist of cell states observed during differentiation from SCC-1 to SCC-2. \textbf{J} The state space shown in (G) without pruning of cell states and cell state transitions. \textbf{K} Morphologies 3, 4 and 5 from the main text are shown at the end of their respective developments (12,000 DTS), along with their simplified state spaces. The layout of the cell states in the morphology mirrors the layout of the cell states in the state space. The green box in the morphology-5 state space corresponds to an SCC disconnected from the other two (i.e., no unidirectional transitions).}
\label{statespace}

\clearpage

\begin{figure}[t]
    \centerfloat
    \includegraphics{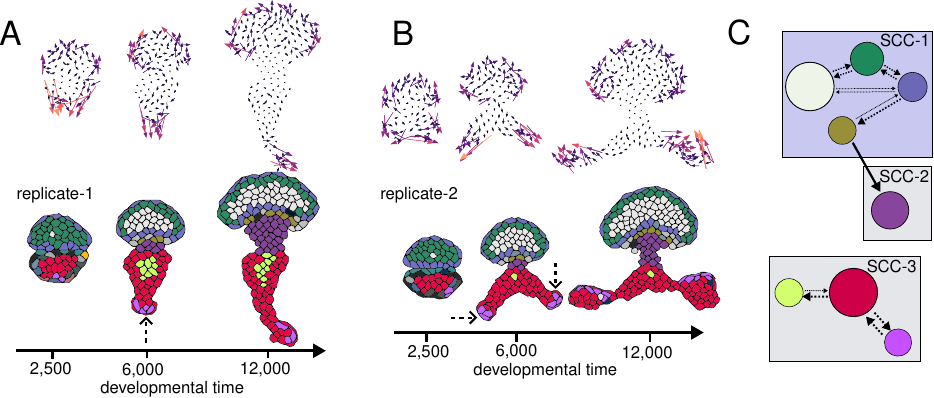}
    \caption{\textbf{Morphogenesis of a morphology with intermediate reproducibility AB}. Developmental replicates at three time points. The vector plots show the displacement of the centre of mass of each cell over the previous 2,000 DTS at each respective time point, with colours indicating magnitude (the lighter, the larger). \textbf{C} Simplified state space of this intermediately reproducible morphology. The state space shows one progenitor-cell type (SCC-1), one differentiated-cell type (SCC-2) and one SCC (SCC-3) that is not part of the progenitor-cell system. The regions of the morphologies corresponding to the progenitor and differentiated cell types appear morphologically similar between developmental replicates. In contrast, the region corresponding to cells in SCC-3, which is not part of the progenitor-cell system, appear dissimilar between developmental replicates. This dissimilarity arises because there is a bifurcation in the motion of these cells in replicate 2 but not in replicate 1 (dashed black arrows). This bifurcation occurs because both moving and stationary states are in SCC-3, as the vector plots confirm. Thus, this morphologies' development shows properties of both highly (SCC-1 and SCC-2) and poorly (SCC-3) reproducible morphogenesis.}
\end{figure}

\clearpage

\begin{figure}
    \centerfloat
    \includegraphics{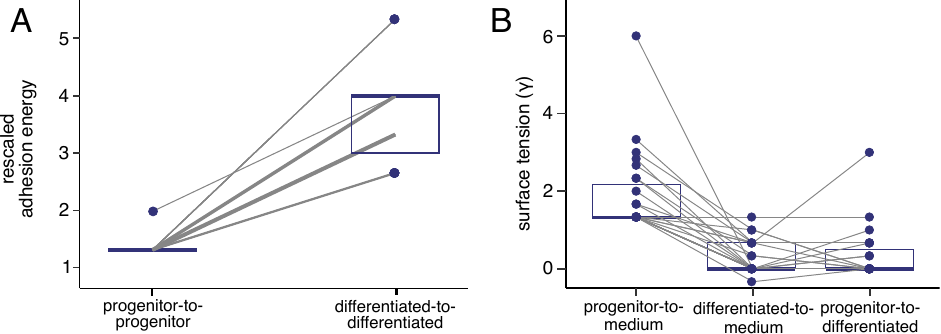}    
\end{figure}
\captionof{figure}{\textbf{Adhesive properties of progenitor-cell systems. A} Adhesion energy arising from progenitor-to-progenitor cell contacts, progenitor-to-differentiated cell contacts and differentiated-to-differentiated cell contacts across the 30 progenitor-cell systems from the 24 morphologies with progenitor-cell systems. The grey lines connect adhesion energies from the same progenitor-cell system. Thicker grey lines indicate multiple progenitor-cell systems with the same data. Adhesion energies are the $J_{ij}$ values arising from contact between the most frequently observed state of each cell type to itself (progenitor-to-progenitor or differentiated-to-differentiated) or the most frequently observed progenitor-cell state to its corresponding most frequently observed differentiated-cell state (progenitor-to-differentiated). Boxes show medians and interquartile ranges across the 30 progenitor-cell systems (The 25th and 75th percentiles are equal to the median for progenitor-to-progenitor and differentiated-to-differentiated adhesion energies; the 75th percentile is equal to the median for differentiated-to-differentiated adhesion energies). The \textit{y}-axis plots the adhesion energies divided by the temperature-like parameter $T$ (set to $T=3$ for all simulations), since this parameter rescales all energy barriers in the CPM. \textbf{B} Progenitor-to-medium surface tension ($n=30$), progenitor-to-differentiated surface tension ($n=30$) and differentiated-to-medium surface tension ($n=26$) across the 24 morphologies that evolved progenitor-cell systems in main text simulations. The median equals the 25th percentile (bottom of the box) because most progenitor-cell types have $\gamma=1.33$, and most differentiated-cell types have $\gamma=0$. See Methods 4.4 for an explanation of surface tensions. }

\begin{figure}[t]
    \centerfloat
    \includegraphics{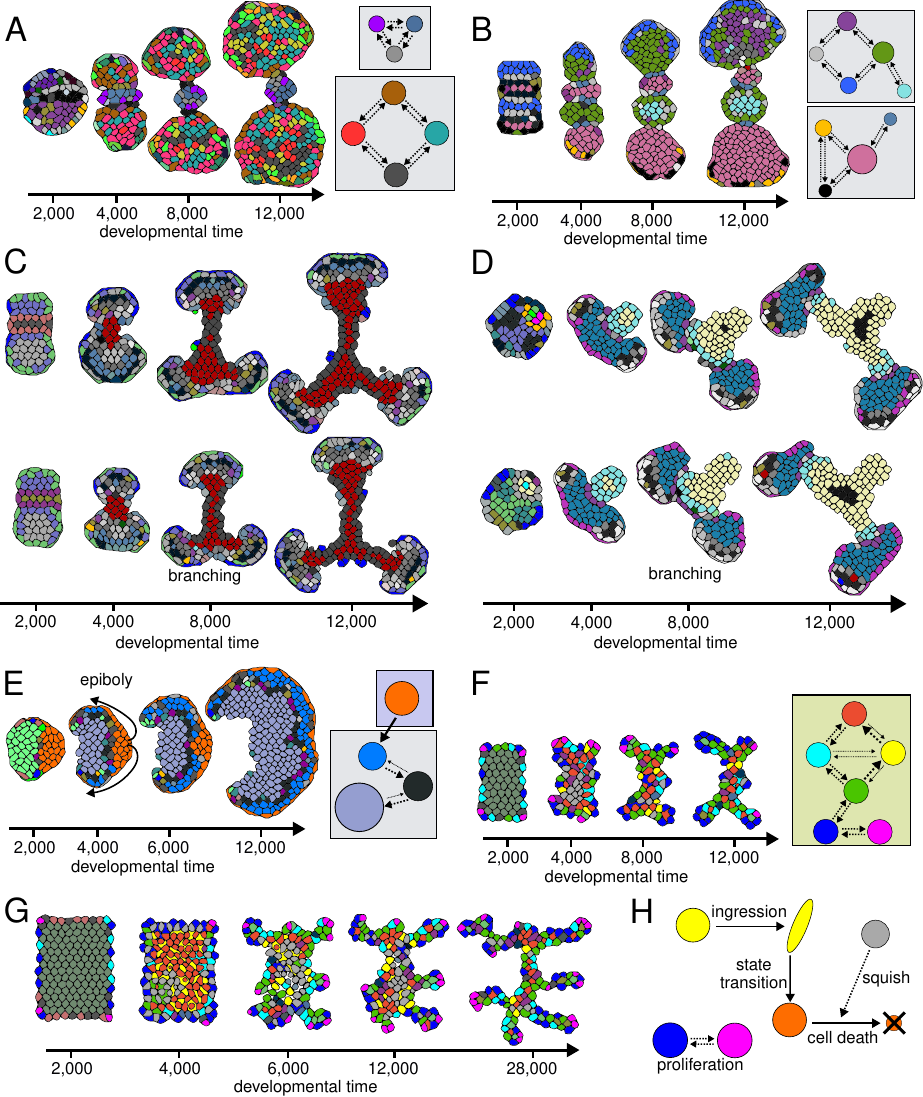}
\end{figure}
\begin{figure}[t]
    \centering
    \includegraphics{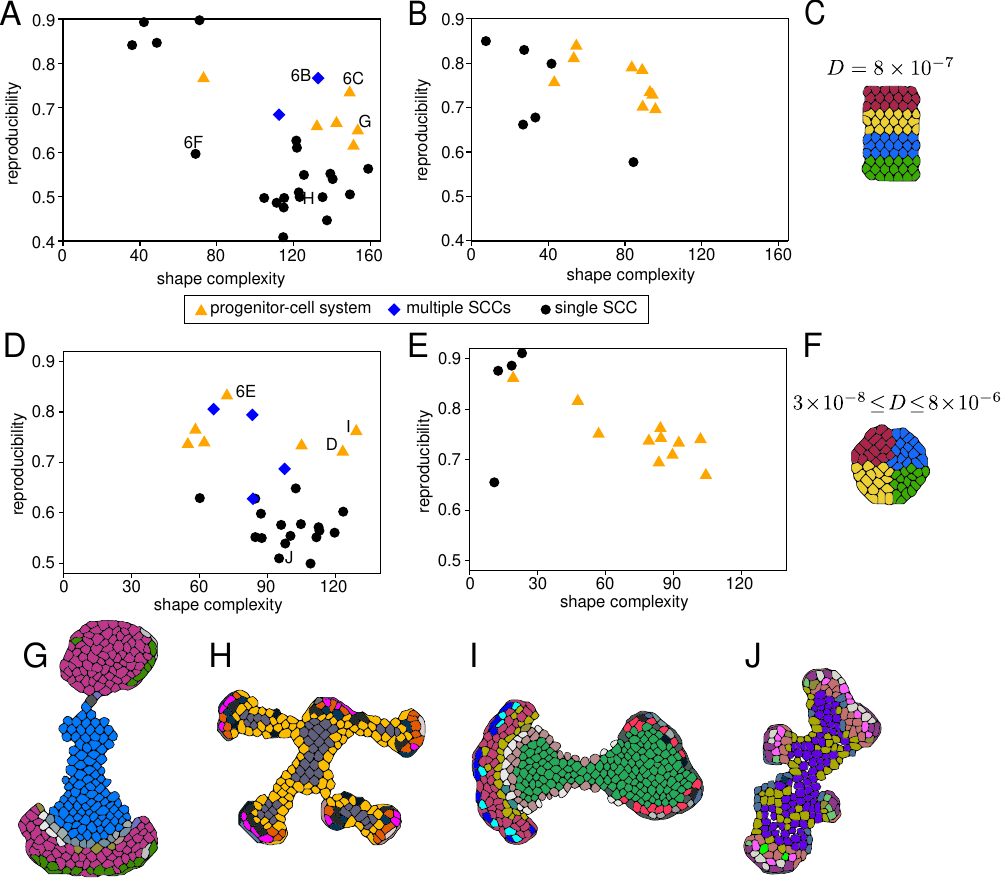}
    \caption{Evolution of progenitor-cell systems in different kinds of simulations. \textbf{AB} Reproducibility and shape complexity of evolved morphologies from simulations where initial condition is a rectangle instead of a circle. Orange triangles are morphologies that evolved progenitor-cell systems. Blue diamonds are morphologies that have multiple SCCs but not progenitor-cell systems. Black circles are morphologies with a single SCC. Shape complexity is averaged over sixty developmental replicates. The selection pressure for simulations shown in (A) is shape complexity, whereas it is both shape complexity and directional motion in (B), detailed in Methods 4.5. \textbf{C} Initial shape and cell states of morphologies used for the simulations shown in (A) and (B). The value $D$ is the diffusivity of morphogens, which is constant for all simulations at $8\times10^{-7}$. \textbf{DE} Reproducibility and shape complexity of evolved morphologies from simulations where morphogen diffusivity mutates. Colour and shape coding of data points is the same as (A) and (B). The selection pressure for simulations shown in (C) is shape complexity, whereas it is both shape complexity and directional motion in (D). See Methods 4.5 for a description of how morphogen diffusivity mutates. \textbf{F} Initial shape and cell states of morphologies used for the simulations shown in (C) and (D). The morphogen diffusivity mutates over the range $3\times10^{-8} \leq D \leq 8\times10^{-6} $. \textbf{GH} Two evolved morphologies after 12,000 DTS from simulations shown in (A). (G) has a progenitor-cell system; (H) does not. \textbf{IJ} Two evolved morphologies after 12,000 DTS from simulations shown in (D). (I) has a progenitor-cell system; (J) does not. The numbers and types of genes used for all the results presented in this figure were: $N_{morph}=4$, $N_{TF}=10$, $N_{tens}=2$, $N_{pairs}=4$ and $N_{med}=3$. }
    \label{fig8}
\end{figure}

\clearpage
\captionof{figure}{\textbf{Diversity of developmental dynamics in our evolved model. AB} Development of two morphologies categorised as having both a complex and reproducible morphology that do not have progenitor-cell systems, each shown at four DTS during their development. Their states spaces show cell states partitioned into two strongly connected components (SCCs, grey boxes) with no unidirectional transitions. The morphology undergoes morphogenesis by stacking ``balls'' of cells on top of each other, with differential adhesion between cells from different balls keeping each ball separated. Both morphologies are highly reproducible because all cells are classed as moving, i.e., there are no stationary cells. The morphology in (A) is indicated by a blue diamond in Fig.\,3C. The morphology is (B) is indicated by a blue diamond in Fig.\,\ref{fig8}A. \textbf{CD} Development of two morphologies with progenitor-cell systems that undergo branching, each shown at four DTS during their development. We categorise this type of morphogenesis as branching instead of a bifurcation because it occurs consistently over developmental replicates. Branching occurs by progenitor-cell differentiation at the tip, which splits the group of progenitor cells in two. \textbf{E} Development of a morphology with a progenitor-cell system that undergoes epiboly shown at four DTS during its development. The orange progenitor cells spread out over the surface of the morphology before differentiating at the top and bottom edge. The morphology is highly reproducible (Fig.\,\ref{fig8}D) \textbf{F} Development of a morphology where cells die to produce protrusions, shown at four DTS during its development. Its development is shown alongside a state space that shows it has only a single SCC. \textbf{G} Development of the same morphology shown in (F) except with three times as many initial cells, shown at five DTS during its development (with development extended to 28,000 DTS). The development shows that the protrusions have a characteristic width of 2-3 cells wide. \textbf{H} Schematic indicating the development dynamics of the morphology shown in (FG). When the width of the morphology is larger than the characteristic width, cells in the yellow state ingress into the centre where they transition to orange cells. The orange cells are squished by the grey cells and die. The blue cells proliferate to extend the protrusions. The morphology is poorly reproducible (see Fig.\,\ref{fig8}D).}
\label{fig7}
\clearpage

\clearpage


\begin{figure}[t!]
    \centerfloat
    \includegraphics{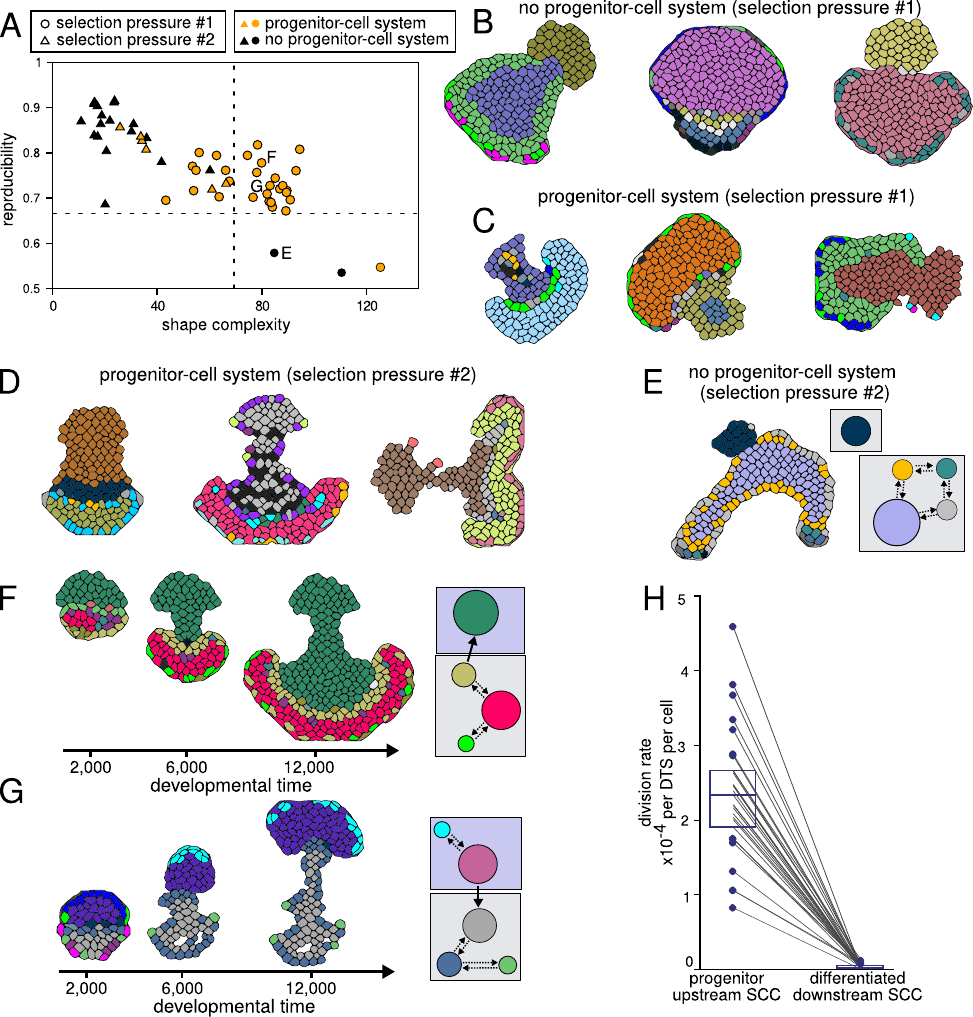}
\end{figure}
\clearpage
    \captionof{figure}{\textbf{Reproducible morphogenesis with progenitor-cell systems evolves under different selection pressures. A} Reproducibility scores and shape complexity of morphologies evolved under two alternative fitness criteria. Fitness criterion \#1 is how much a morphology shifts its centre of mass over the 12,000 DTS (Methods 4.5). Fitness criterion \#2 is a sum of the shift in the centre of mass criterion and the shape complexity criterion (Methods 4.5). Triangles are shape complexity (not fitness) and reproducibility scores of evolved morphologies from 25 simulations using fitness criterion \#1. Circles are shape complexity and reproducibility scores of evolved morphologies from 31 simulations using fitness criterion \#2. Data colour-coded orange indicates morphologies with progenitor-cell systems. Data colour-coded black indicates morphologies without progenitor-cell systems. The dashed line is a shape complexity of 70, the threshold we used to determine if a morphology was sufficiently morphologically complex in the original set of simulations, where the selection pressure was only shape complexity (described in Methods 4.5). The dashed line at a reproducibility score of 66\% is the cut-off for high reproducibility indicated in Figure 3A in the main text. \textbf{B} Three evolved morphologies after 12,000 DTS from simulations using fitness criterion \#1 that did not evolve progenitor-cell systems. \textbf{C} Three evolved morphologies after 12,000 DTS from simulations using fitness criterion \#1 that did evolve progenitor-cell systems. \textbf{D} Three evolved morphologies after 12,000 DTS from simulations using fitness criterion \#2 that did evolve progenitor-cell systems. \textbf{E} One of the two evolved morphologies without a progenitor-cell system from the simulations using fitness criterion \#2, shown after 12,000 DTS. Its state space is shown to the right. Although the state space shows multiple SCCs, there is no unidirectional transition between them and thus no progenitor-cell system. The morphology is poorly reproducible because one of its SCCs has both moving and stationary states in it. \textbf{FG} Development of two morphologies with progenitor-cell systems. Morphologies are shown after 2,000, 6,000 and 12,000 DTS. Simplified cell state spaces are shown to the right, with these cell state spaces indicating multiple SCCs with unidirectional transitions. \textbf{H} The rate at which cells divide per developmental time when their state belongs to an upstream SCC (left) or a downstream SCC (right). Each data point represents an SCC from the 29 evolved morphologies that had multiple SCCs with unidirectional transitions that were evolved under selection for directional motion and shape complexity. Black lines connect upstream SCCs to their counterpart downstream SCCs. Boxes show medians and interquartile ranges. The number and types of genes for simulations presented in this Figure is the same as the main text ($N_{morph}=3$, $N_{TF}=9$, $N_{tens}=2$, $N_{pairs}=5$ and $N_{med}=5$.) }
    \label{asym}

\clearpage

\begin{figure}[t!]
    \centerfloat
    \includegraphics{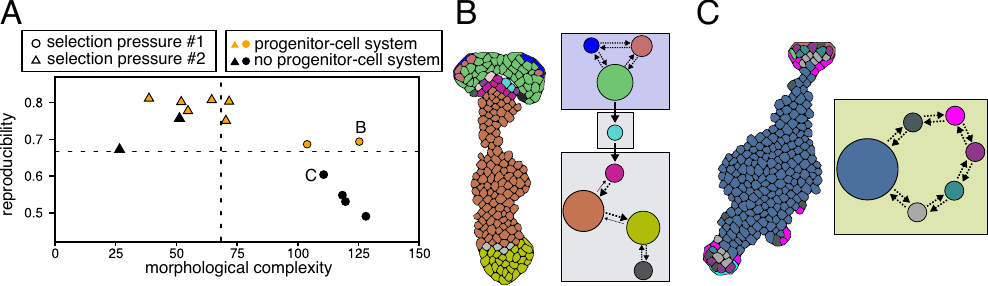}    
    \caption{\textbf{Evolution of progenitor-cell systems and high reproducibility with a minimal genome. A} Reproducibility scores and shape complexity of morphologies evolved under two alternative fitness criteria using a minimal genome model. Fitness criterion \#1 is how much a morphology shifts its centre of mass over the 12,000 DTS (Methods 4.5). Fitness criterion \#2 is a sum of the shift in the centre of mass criterion and the shape complexity criterion (Methods 4.5). Triangles are shape complexity (not fitness) and reproducibility scores of evolved morphologies from eight simulations using fitness criterion \#1. Circles are shape complexity and reproducibility scores of evolved morphologies from seven simulations using fitness criterion \#2. Data colour-coded orange indicates morphologies with progenitor-cell systems. Data colour-coded black indicates morphologies without progenitor-cell systems. The dashed line is a shape complexity of 70, the threshold we used to determine if a morphology was sufficiently morphologically complex in the original set of simulations, where the selection pressure was only shape complexity (described in Methods 4.5). The dashed line at a reproducibility score of 66\% is the cut-off for high reproducibility indicated in Figure 3A in the main text. The minimal genome comprises 12 genes (as opposed to 26): $N_{morph}=3$, $N_{TF}=6$, $N_{tens}=0$, $N_{pairs}=2$ and $N_{med}=2$. The adhesion parameters that we changed to accommodate the smaller genome are $J_{ij}^{max}=20$, $\phi^{ij}_k =4$, $J_{im}^{max}=18$, $\psi_1=10$ and $\psi_2=2$. There are no membrane tension proteins. \textbf{B} A highly reproducible morphology developed for 12,000 DTS from the minimal genome model with a progenitor-cell system. Its state space to the right shows multiple SCCs connected by a unidirectional transition. \textbf{C} A poorly reproducible morphology developed for 12,000 DTS from the minimal genome model without a progenitor-cell system. Its state space to the right shows a single SCC. }   
    \label{genomes}
\end{figure}

\clearpage

\begin{figure}
    \centerfloat
    \includegraphics{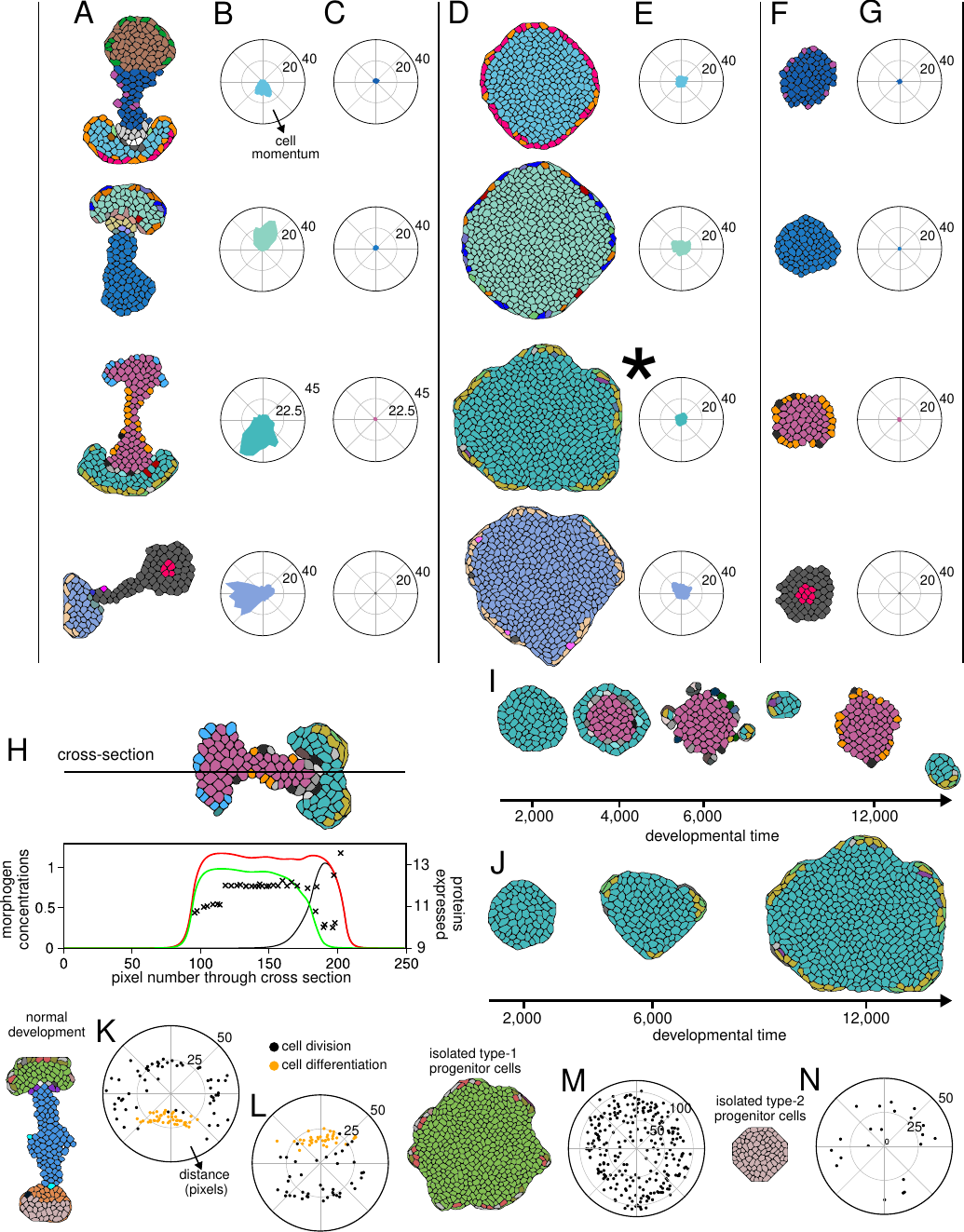}    
\end{figure}
\clearpage
\captionof{figure}{\textbf{Progenitor-cell motion and divisions are isotropic when isolated from other cell types. A} Four evolved morphologies with progenitor-cell systems at their developmental endpoints (12,000 DTS). Each morphology has one progenitor-cell type and one differentiated-cell type \textbf{BC} Polar plots of momentum magnitude by angle of momentum over normal development of the four morphologies, separated by cell type; (B) shows plots for the four progenitor-cell types, where momentum appears anisotropic, and (C) shows plots for the four differentiated-cell types. \textbf{D} Isolated progenitor cells for each morphology at their developmental endpoints (12,000 DTS). \textbf{E} Polar plots of momentum magnitude by the angle of momentum of the isolated progenitor cells for each morphology, showing radially symmetrically distributed motion. \textbf{F} Isolated differentiated cells for each morphology at their developmental endpoints (12,000 DTS). \textbf{G} Polar plots of momentum magnitude by angle of motion of the isolated differentiated cells for each morphology. \textbf{HIJ} Analysis of progenitor-cell differentiation in morphology-1. We show morphology-1 here because its progenitor-cell type is one of the four (out of 30) that differentiates when isolated from other cell types. (H) shows morphogen concentrations along a cross-section of morphology-1 at 8,000 DTS. The sum of cell protein concentrations is also shown for cells along this cross-section (each cross is one cell). The green morphogen induces progenitor-cell differentiation, which is produced by differentiated cells (data not shown). However, progenitor cells also begin to express the green morphogen themselves if the black and red morphogen concentrations get too high (data not shown). (I) shows the development of isolated type-1 progenitor cells after 2000, 4000, 6000 and 12,000 DTS. The isolated progenitor cells differentiate around the cluster's centre around 4,000 DTS. Differentiation begins at the cluster's centre because black and red morphogens are at their highest concentration, resulting in the expression of the green morphogen that induces differentiation. The differentiation of progenitor cells in this isolated morphology causes progenitor and differentiated cells to split apart due to differential adhesion between progenitor and differentiated cells. \textbf{J} Development of morphology-1 progenitor cells in isolation, except with morphogen concentrations prevented from increasing above 1.0. Preventing black and red morphogen concentrations from increasing above 1.0 stops the green morphogen from being expressed and thus prevents progenitor-cell differentiation. For the four progenitor-cell types that autonomously differentiate, including this one, we measured the motion anisotropy of isolated progenitor cells (plotted in Fig.\,5F) after placing restrictions on morphogen concentrations to prevent differentiation. The asterisk in (DE) shows the morphology (D) and polar plot (E) of isolated progenitor-cell motion from morphology-1 when these morphogen restrictions are in place. \textbf{KL} Spatial distribution of cell divisions (black dots) and differentiations (orange dots) for (C) type-1 and (D) type-2 progenitor-cell domains from morphology-6, with the pole of the plots defined as the centre of mass of all of the respective progenitor cells in that cluster (i.e., the pole shifts as progenitor cells move over development). The location of a cell division is marked at the centre of mass of the parent cell. The plots show that type-1 progenitor cell divisions are asymmetric in that they occur around the cluster's periphery rather than the centre. In contrast, type-2 progenitor cell divisions predominantly occur distal to the location of progenitor-cell differentiation. \textbf{MN} Spatial distribution of cell divisions for type-1 and type-2 progenitor cells from morphology-6, as in (KL). }
\label{isotropy}

\begin{figure}
    \centerfloat
    \includegraphics{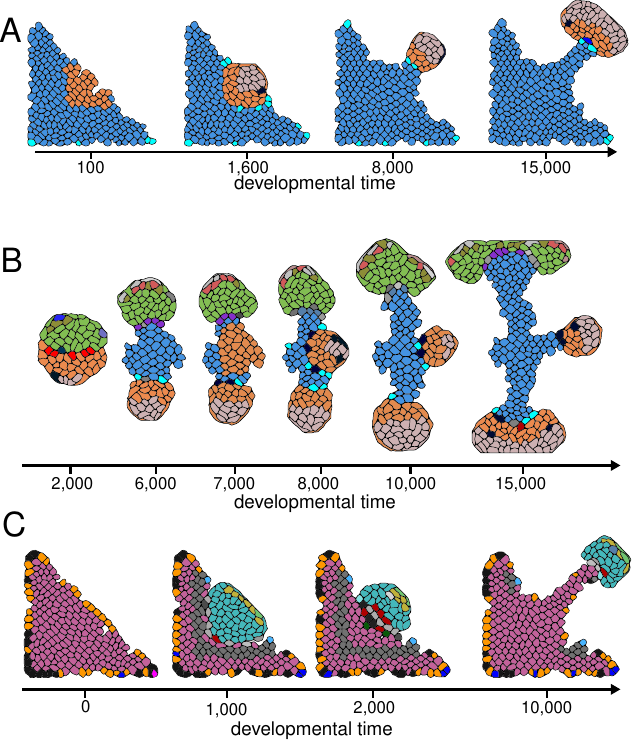}    
\end{figure}
\captionof{figure}{\textbf{Programmability of progenitor-cell-based morphogenesis.} We created arbitrary morphologies of progenitor and differentiated cells from morphology-1 and morphology-6 and simulated their development for morphology-1 and morphology-6. \textbf{A} Development of an arbitrary morphology that begins as a triangular shape of differentiated cells from morphology-6 accompanied by a small domain of type-2 progenitor cells from that morphology on the diagonal surface, with development extended to 15,000 DTS. We programmed this morphology by artificially changing the state of the initial cells to either differentiated cells or progenitor cells depending on their initial position in the triangle. \textbf{B} The same developmental trajectory of morphology-6 as shown in Figure 5A of the main text, except with the state of the differentiated cells on the centre-right flank artificially changed to type-2 progenitor cells at 7,000 DTS. This artificial state-change causes a new branch to form perpendicular to the native one. \textbf{C} An arbitrary morphology that begins as a triangular shape of differentiated cells form morphology-6 accompanied by a small domain of progenitor cells from that morphology on the diagonal surface, with development lasting for 10,000 DTS.}
\label{adhesion}

\clearpage

\begin{figure}
    \centerfloat
    \includegraphics{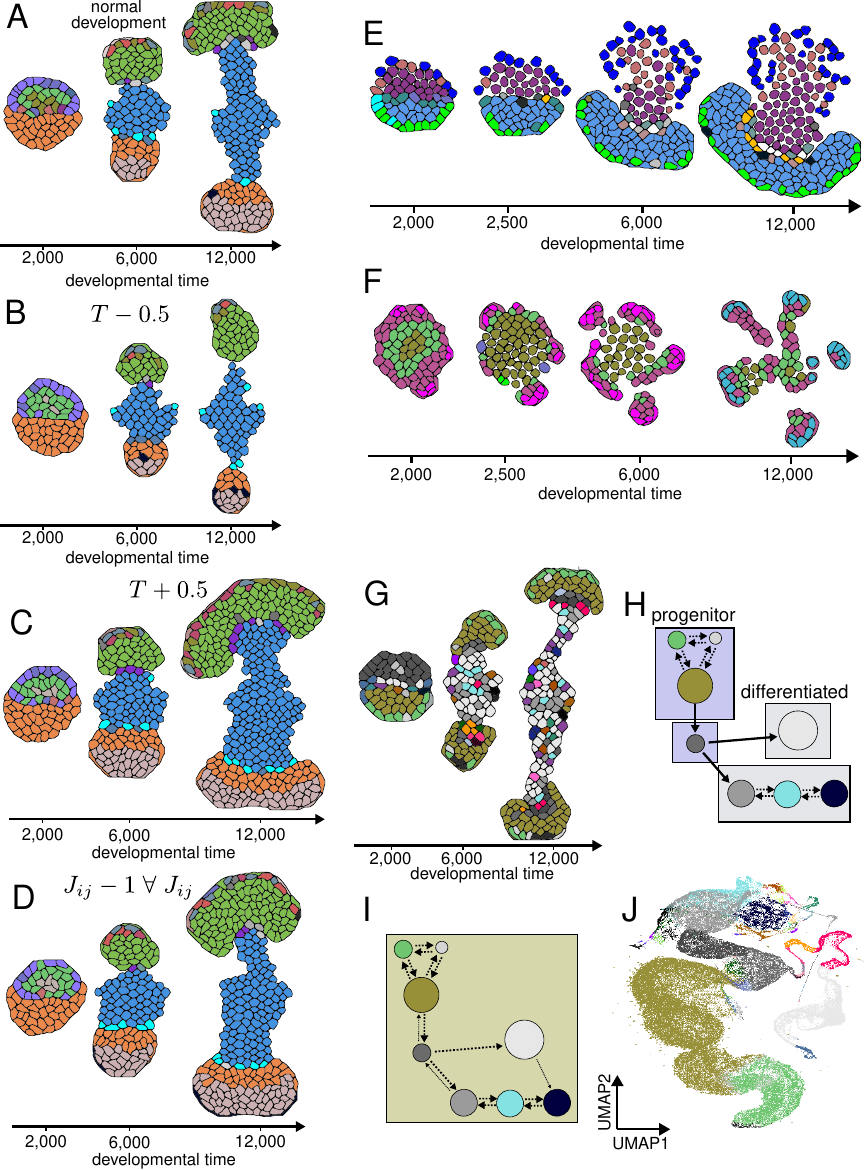}    
\end{figure}
\clearpage
\captionof{figure}{\textbf{Technical details of the model and analysis. ABCD} Robustness of morphology-6 morphogenesis to changes in temperature and adhesion parameters. (A) shows the normal development of morphology-6 after 2000, 6000 and 12,000 DTS. (B) shows the development of morphology-6 using the same random seed as (A) but with the temperature, $T$, decreased by 0.5 (where $T=3$ by default). (C) shows the same development of morphology-6 using the same random seed as (A) but with $T$ increased by 0.5. (D) shows the same development of morphology-6 using the same random seed as (A) but with all cell-cell adhesion energies, $J_{ij}$, decreased by one. \textbf{EF} Morphologies that are designated 0 fitness because there is no single shape to measure the complexity of. We show the development of two morphologies at four DTS. The cells of both morphologies split up from each other during development. These morphologies were arbitrarily chosen from evolving populations in evolutionary simulations. \textbf{GHIJ} One of the two morphologies with progenitor-cell systems that has rare transitions from differentiated to progenitor cells (these rare transitions were removed during pruning). (G) shows a developmental replicate of the morphology, and (H) shows its state space after pruning, indicating multiple SCCs with unidirectional transitions. (I) shows the same state space before the pruning of rare transitions (although rare cell states are pruned for clarity), indicating only a single SCC. (J) is a visualisation of cell states (colours) mapped onto cell protein expression profiles (points) that have undergone dimension reduction by UMAP for the morphology shown in (G). Profiles are collected from four developmental replicates. Two lines in the UMAP plot connect the white cell state (deemed to be differentiated in the pruned state space) to other cell states. The thinner of the two lines corresponds to a transition of this differentiated cell state to the other differentiated cell state. This transition is one of those removed after pruning.}
\label{methods}

\clearpage

\begin{figure}[t]
    \centerfloat
    \includegraphics{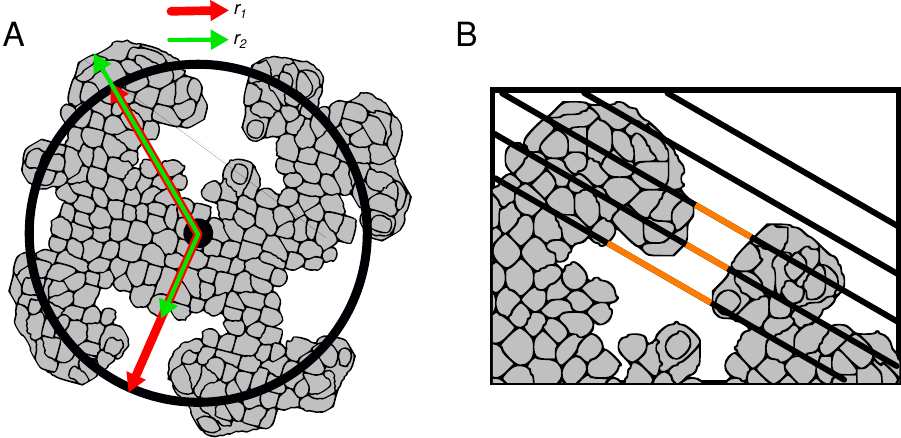}
    \caption{\textbf{Visual illustration of algorithm used to quantify shape complexity. A} Visualization of the measure of deviation from a circle. We show the difference between the actual morphology radius ($r_1$) and the radius if all morphology pixels were to be circularly distributed ($r_2$, black circle) at two locations. \textbf{B} Visualization of the measure of inward folds. Solid black lines depict the evenly spaces parallel lines drawn across the grid at one of the 12 evenly-spaced angles. The actual number of lines occur with a much higher density than the five shown here. The orange segments of the lines indicate ``gaps'' in the morphology. The square root of the total number of these gaps with a minimum length of 20 pixels is the morphologies' negative curvature.}
    \label{illustration}
\end{figure}

\clearpage

\bibliography{bibliography}
\bibliographystyle{unsrt}